\documentclass[preprint,aps,12pt]{revtex4}

\usepackage{graphicx}
\usepackage{dcolumn}
\usepackage{bm}
\usepackage{color}

\usepackage{amsmath}
\usepackage{amssymb}

\begin{document}

\preprint{T04/073}
\preprint{TUW-04-13}

\title{Isolating vacuum amplitudes in quantum field calculations
at finite temperature}

\author{Jean-Paul Blaizot}
\email{blaizot@spht.saclay.cea.fr}
\affiliation{Service de Physique Th\'eorique, CEA/DSM/SPhT,
91191 Gif-sur-Yvette Cedex, France.}
\author{Urko Reinosa}
\email{reinosa@hep.itp.tuwien.ac.at}
\affiliation{Institut f\"ur Theoretische Physik, Technische Univerist\"at Wien, Wiedner Hauptstrasse 8-10/136, A-1040 Wien Austria.}
\date{\today}

\begin{abstract}
In calculating Feynman diagrams at finite temperature, it is
sometimes convenient to
   isolate subdiagrams  which  do not depend explicitly on the
temperature. We show that, in the imaginary
time formalism, such a separation can be achieved easily by
exploiting a simple method, due to M. Gaudin,
to perform the sum over the Matsubara frequencies. In order to 
manipulate freely contributions which may be individually
singular, a regularization has to be introduced. We show that, in 
some cases,  it is possible to choose this regularization in
such a way that the isolated subdiagrams can be identified with 
analytical continuations
of vacuum $n$-point functions.
  As an aside illustration of Gaudin's method, we use it to
  prove the main part of a recent conjecture concerning the relation
  which exists in  the imaginary time formalism between
the expressions of a Feynman diagram at zero and finite temperature.
\end{abstract}

\pacs{Valid PACS appear here}
\maketitle

\newcommand \beq{\begin{eqnarray}}
\newcommand \eeq{\end{eqnarray}}
\newcommand \bea{\begin{eqnarray}}
\newcommand \eea{\end{eqnarray}}
\newcommand \ga{\raisebox{-.5ex}{$\stackrel{>}{\sim}$}}
\newcommand \la{\raisebox{-.5ex}{$\stackrel{<}{\sim}$}}
\def\simge{\mathrel{%
      \rlap{\raise 0.511ex \hbox{$>$}}{\lower 0.511ex \hbox{$\sim$}}}}
\def\simle{\mathrel{
      \rlap{\raise 0.511ex \hbox{$<$}}{\lower 0.511ex \hbox{$\sim$}}}}

\def\simle{\mathrel{
      \rlap{\raise 0.511ex \hbox{$<$}}{\lower 0.511ex \hbox{$\sim$}}}}
\oddsidemargin=8pt
\parindent=10pt

\newtheorem{lem}{Lemme}
\newtheorem{prop}{Proposition}
\newenvironment{dem}{\noindent{\bf
D\'emonstration\\}}{\flushright$\diamond$\\}
\newtheorem{rem}{Remarque}

\section{Introduction}

In doing quantum field theory calculations at finite temperature 
\cite{Kapusta:tk,LeBellac96}, it is often useful to separate
contributions of subdiagrams which do not explicitly depend on the 
temperature. In some cases,
these subdiagrams can
be    identified  with
amplitudes calculated with the rules of zero temperature, and we 
shall refer to them  as vacuum
amplitudes. Isolating such amplitudes is useful in particular  to 
analyze the ultraviolet divergences.
These divergences are associated with the short distance
singularities in the propagator, and these are not modified by the 
temperature \cite{LeBellac96,Collins:xc}. Thus, one expects the
ultraviolet divergences to be those of the vacuum subdiagrams.

Identifying vacuum amplitudes in an arbitrary Feynman diagram at 
finite temperature is, in general, not an
easy task. In the imaginary time formalism, which is the one we shall 
use in this paper,  the
calculation of a Feynman diagram at finite temperature is similar to 
the corresponding calculation at zero temperature, the
integrals over energies being replaced by sums over the discrete 
Matsubara frequencies. (Note also
that the sums over Matsubara frequencies go over the Euclidean
integrals of field theory in the limit of zero temperature.) In this 
formalism, the temperature dependence appears explicitly,
after the sums over Matsubara frequencies have been done, through 
statistical factors which vanish
when the temperature vanishes. This makes it easy, in principle, to 
identify the zero temperature contributions. This procedure
works well for one loop-diagrams which are linear in the statistical 
factors, but things become  more subtle in higher loop orders.
There is indeed a further complication. Depending on how one proceeds 
to perform the calculation, one may end up
with expressions containing different numbers of statistical factors, 
or statistical factors whose arguments
involve energies attached to different lines of the diagram. This 
makes the separation of various contributions difficult.

There is
however a method to perform the sums over Matsubara frequencies which 
leads directly to a result where the number of statistical
factors is   related to the number of loops, and the arguments of the 
statistical factors are energies attached to single
lines of the diagram. This method was developed by M. Gaudin a long 
time ago \cite{Gaudin65}, but seems to have been largely
ignored in the recent literature. We shall use   this method to 
analyze the separation of vacuum and thermal contributions in a
general Feynman diagram.
The main problem in doing the sum over Matsubara frequencies is the 
choice of the independent frequencies. In his work,  Gaudin
   relates the various choices of
independent variables to the various trees that one can build with
the lines of the diagram \cite{Gaudin65}.   Note that, more recently 
\cite{Guerin:ny}, tree diagrams were also used in a
similar context,  but the   general rules that apply to arbitrary 
loop order were not given.

We shall show that the rules  proposed by Gaudin allow us to organize 
the result of a
calculation of a Feynman diagram at finite temperature according to powers of
statistical factors. In some cases, this will allow us  to isolate 
vacuum amplitudes, that is vacuum subdiagrams that one can
relate  to analytic continuations of
Euclidean vacuum amplitudes. This connection is not always possible 
to realize however. This is because, at some intermediate
step of the analysis, one needs to introduce a regularization to give 
meaning to otherwise singular individual contributions
(whose sum is regular). For the regularization that we use, we can 
find counter-examples of vacuum subdiagrams which are not analytic 
continuations
of corresponding, i.e., topologically identical, vacuum Euclidean amplitudes.

The paper is organized as follows. In the next section, we consider 
several examples which illustrate various features of
calculations of Feynman diagrams at finite temperature. These 
examples allow us to specify concretely what is involved in 
identifying vacuum
amplitudes. They also serve as a pedagogical introduction for the 
general method of summing over Matsubara frequencies that is
discussed in Sect.~\ref{rules}. The rules presented in 
Sect.~\ref{rules} are essentially those derived by M. Gaudin 
\cite{Gaudin65}. They are used in
Sect.~\ref{section:thermal} in order to write down an expansion of 
the   temperature dependent pieces of a Feynman diagram in terms of 
powers
of statistical factors (which vanish as the temperature vanishes). We 
show that this decomposition exists only if some
regularization is introduced to control individually singular terms. 
In some cases we can relate the vacuum subdiagrams to well
defined analytic continuations of Euclidean amplitudes. But it is not 
always possible to do so, as illustrated by a
counter-example that we present at the end of 
Sect.~\ref{section:thermal}. Conclusions are presented in 
Sect.~\ref{sec:conclu}.
Finally, in  the appendix, we show how the main part of  the 
conjecture of Ref.~\cite{Esp:2003} follows directly
form the rules of Sect.~\ref{rules}.

\section{Simple examples}
In this section, we work out simple examples of finite
temperature calculations in scalar field
theory. Our goal, beyond introducing the basic notation, is to
illustrate on a few cases how one can
isolate, in a given Feynman diagram,   a  contribution to either the
full diagram  or to a subdiagram,
which does not explicitly depend on the temperature. In the cases
studied here, such contributions can be  associated to  analytic 
continuations of vacuum
$n$-point functions. As we proceed, we shall also
recall some elementary techniques to perform sums over Matsubara
frequencies which will be generalized in
the next section.

\subsection{General definitions}
In the imaginary time formalism, the (time-ordered) propagator of a 
free scalar field can be  written
in a mixed representation, as a function of
momentum  ${\bf p}$ and imaginary time $\tau$, as follows \cite{LeBellac96}:
\beq\label{taupropag}
D_0(\tau_1-\tau_2,{\bf p})&=&\int {\rm d}^3 x \,\, {\rm e}^{-i{\bf
p}\cdot({\bf x}_1-{\bf x}_2)}\,\,\langle {\rm T }
\phi(\tau_1,{\bf x}_1)\phi(\tau_2,{\bf x}_2)\rangle\nonumber\\
   &=&\frac{1}{2\varepsilon_p}\left\{
(1+n_{\varepsilon_p})e^{-\varepsilon_p|\tau_1-\tau_2|}
+n_{\varepsilon_p}e^{\varepsilon_p|\tau_1-\tau_2|}\right\}.
\eeq
In this expression, valid  for $|\tau_1-\tau_2|\leq
\beta=1/T$, with $T$ the temperature, $n_{\varepsilon_p}$ is the
Bose-Einstein statistical factor:
\beq
n_\varepsilon=\frac{1}{{\rm e}^{\beta\varepsilon}-1},\qquad
n_{-\varepsilon}=-1-n_{\varepsilon},
\eeq
and $\varepsilon_p=\sqrt{p^2+m^2}$ is the energy of a mode with
momentum $p$. We shall write the
statistical factor indifferently as $n_\varepsilon$ or
$n(\varepsilon)$. The  propagator in
Eq.~(\ref{taupropag}) can be used in perturbative calculations at finite
temperature. The calculation of
a diagram $\Gamma$ proceeds then typically as follows: After having
chosen an orientation
\footnote{In the present case the orientation is completely
arbitrary. In theories with a
conserved charge, like in fermionic theories, one choses the
orientation so as to conserve the
flow of charges at each vertex.} for each line of
$\Gamma$,
   one associates to each vertex $v_i$ of $\Gamma$ an  imaginary
time $\tau_i$, and to a line joining the vertex $v_i$ to the vertex
$v_j$ one
    associates the propagator $D_0(\tau_j-\tau_i,{\bf p})$. The contribution
of the diagram is then obtained by integrating over all the time
variables $\tau_i$ between 0 and $\beta$.
We shall give soon an  example of such a calculation. For simple diagrams, this
technique can be quite convenient, and indeed it has been used in a
systematic analysis of  the one loop
contributions in QCD at finite temperature \cite{Pisarski:1987wc}.
However, because the
propagators take different forms according to the sign of
$\tau_j-\tau_i$, one needs to treat separately the various
integration subdomains, and this  becomes
rapidly cumbersome for high order diagrams.

An alternative is to use an energy (or frequency) representation
  of the propagator,
which we shall write in the form ($\omega$ is a complex variable):
\beq\label{represspectrale}
D(\omega,{\bf p})=\int_{-\infty}^{+\infty}\frac{{\rm
d}p_0}{2\pi}\frac{\rho(p_0,{\bf p})}{p_0-\omega},
\eeq
where the quantity $\rho(p_0,{\bf p})$ is
the spectral function. It is an odd function of $p_0$. In most of the
arguments of this paper, the spectral function needs not be that of 
free particles.  We
call  $\rho_0$ and $D_0$,
respectively, the spectral function and the propagator for  free
particles:
\beq\label{rho0D0}
\rho_0(p_0,{\bf p})=\frac{\pi}{\varepsilon_p}\left[\delta(p_0-\varepsilon_p)-\delta(p_0+\varepsilon_p)\right],\qquad
D_0(\omega,{\bf p})=\frac{1}{\varepsilon_p^2-\omega^2}.
\eeq

The propagator (\ref{represspectrale}) is an analytic function of
$\omega$ everywhere in the complex $\omega$-plane, except on the real
axis; above (below) the real axis it coincides with   the retarded
(advanced) propagator.  The imaginary time propagator
(\ref{taupropag}) can be recovered from
the Fourier coefficients
$D(i\omega_n,{\bf p})$, where $\omega_n=2\pi nT$ is a Matsubara frequency:
\beq\label{DtauFourier}
D(\tau,{\bf p})=\frac{1}{\beta}\sum_n{\rm e}^{i\omega_n\tau}
D(i\omega_n,{\bf p}).
\eeq
While the sum in Eq.~(\ref{DtauFourier}) can be done easily for free
particles by using the expression of $D_0(\omega,{\bf p})$
given above, Eq.~(\ref{rho0D0}), we shall often perform the sum over
Matsubara frequencies after using the spectral
representation of the propagator, Eq.~(\ref{represspectrale}). Then
the following formula will be useful:
\beq\label{regularsum}
\frac{1}{\beta}\sum_n\frac{{\rm
e}^{i\omega_n\tau}}{p_0-i\omega_n}=\epsilon_\tau\,n({\epsilon_\tau\,p_0})\,{\rm e}^{p_0\tau},
\eeq
where $\epsilon_\tau=1$ if $\tau>0$ and $\epsilon_\tau=-1$ if
$\tau<0$. It follows that:
\beq\label{DtauFourier1}
D(\tau,{\bf p})=\int_{-\infty}^{+\infty}\frac{{\rm
d}p_0}{2\pi} {\rm e}^{p_0\tau}
    \rho(p_0,{\bf p})\,\epsilon_\tau n({\epsilon_\tau p_0}).
\eeq
The r.h.s. of Eq.~(\ref{DtauFourier1}) provides two equivalent ways to
calculate $D(\tau=0,{\bf p})$, by letting $\tau$ go to 0 by positive or
negative values. Both are of course equivalent. (Note that
$D(\tau,{\bf p})$ is not analytic at
$\tau=0$: it is continuous,  but it's derivative is not.)

The evaluation of the fluctuations of the free scalar field provides the
simplest example where the separation of vacuum and thermal
contributions can be realized easily.
Note that since the statistical factors are explicit in the
expression  (\ref{taupropag}) of the
propagator, using this expression makes it
straightforward to separate these contributions.
Indeed, by using  Eq.~(\ref{taupropag}) one obtains immediately:
\beq\label{I0I1}
\langle\phi^2\rangle=\int\frac{d^3p}{(2\pi)^3}D_0(\tau=0,{\bf
p})=\int\frac{d^3p}{(2\pi)^3}\frac{1}{2\varepsilon_p}\left(
1+2n_{\varepsilon_p}\right)\equiv I_0+I_1,
\eeq
where the two terms in the right hand side  are
the zero temperature contribution $I_0$  and the finite temperature one $I_1$.

Alternatively, we can use the Fourier representation, Eq.~(\ref{DtauFourier}),
and Eq.~(\ref{DtauFourier1}), in order to write $D_0(\tau=0,{\bf p})$ as:
\beq\label{thermaldec}
\int_{-\infty}^{+\infty}\frac{{\rm d}p_0}{2\pi} \rho(p_0,{\bf p}) n(p_0)
=-\int_{-\infty}^{+\infty}\frac{{\rm d}p_0}{2\pi}  \rho(p_0,{\bf p})
\theta(-p_0)
+\int_{-\infty}^{+\infty}\frac{{\rm d}p_0}{2\pi}
\rho(p_0,{\bf p})\epsilon(p_0)n({|p_0|}).
\eeq
Now the separation of the thermal contribution is achieved with
the help of the formula:
\begin{equation}\label{statfactor}
n({p_0})=-\theta(-p_0)+\epsilon(p_0)n({|p_0|}),
\end{equation}
where $\epsilon(p_0)=1$ or $\epsilon(p_0)=-1$ depending on the sign of $p_0$
($\epsilon(p_0)=\theta(p_0)-\theta(-p_0))$. As $T\to
0$,
$n({|p_0|})\to 0$ and
$n({p_0})\to -\theta(-p_0)$. Note that  since
$\rho(p_0)$ is an odd function we have:
\beq
-\int_{-\infty}^{+\infty}\frac{{\rm d}p_0}{2\pi}  \rho(p_0,{\bf p})
\theta(-p_0) =
\int_{-\infty}^{+\infty}\frac{{\rm d}p_0}{2\pi}  \rho(p_0,{\bf p})
\theta(p_0).
\eeq
This gives us two possibilities
to write the zero temperature piece in Eq.~(\ref{thermaldec}), either
as $-\theta(-p_0)$ (as we
have done in Eq.~(\ref{thermaldec})), or as
$\theta(p_0)$. These two choices are of course closely related to the
two ways in which one can let
$\tau$ go to zero to get $D(\tau=0,{\bf p})$ from
Eq. (\ref{DtauFourier1}). As for the last term in
Eq.~(\ref{thermaldec}), it is useful to note that it is unaffected by
the change in the sign of
$p_0$.

Because $I_0$ is ultraviolet divergent, it may be convenient to write
it as an Euclidean integral:
\beq\label{I0Euclidean}
I_0=\int\frac{d^4p}{(2\pi)^4} D(ip_0,{\bf p}).
\eeq
This   allows us in particular to use covariant
regulators to evaluate it.

The finite temperature contribution $I_1$ is defined in
Eq.~(\ref{I0I1}). It can  also be obtained from
the last term in Eq.~(\ref{thermaldec}), which allows us to write it
also as a 4-dimensional integral,
which will prove convenient in our forthcoming analysis. We shall
introduce a special notation for the
integrand in this last term of Eq.~(\ref{thermaldec}):
\beq\label{thermalrho}
\sigma(p_0,{\bf p})\equiv \rho(p_0,{\bf p})\epsilon(p_0)n({|p_0|}).
\eeq
This particular combination of the spectral density and the statistical factor
will appear systematically in the calculations. The function
$\sigma(p_0,p)$ is an even, positive, function of $p_0$.  For free
particles, $\sigma=\sigma_0$,
with:
\beq
\sigma_0(p_0,{\bf
p})=\frac{\pi}{\varepsilon_p}\left[\delta(p_0-\varepsilon_p)+\delta(p_0+\varepsilon_p)\right] n_{\varepsilon_p}.
\eeq
    In terms of
$\sigma$ the thermal contribution to $\langle\phi^2\rangle$ is simply:
\beq
I_1=\int\frac{d^4p}{(2\pi)^4} \, \sigma(p_0,{\bf p}).
\eeq

\subsection{One loop in two ways}
We now proceed to the analysis of our first non trivial example, that of
the one loop contribution to the self-energy in a $\phi^3$ scalar
theory. For this example, we shall present two calculations, one
using the time  representation, one using
the frequency representation. In both cases, we shall focus on the
separation of the contribution
which depends on the temperature from that which does not.

\subsubsection{Time representation}

\begin{figure}[htbp]
\begin{center}
\includegraphics[width=7cm]{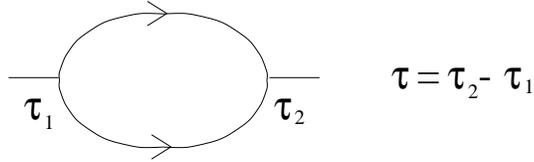}
\caption{One-loop self-energy in $\phi^3$ scalar field theory, time
representation\label{fig:1loop_time}}
\end{center}
\end{figure}

We start by writing the diagram in the time representation (see
Fig.~\ref{fig:1loop_time}). For the particular choice of orientation
given in
Fig.~\ref{fig:1loop_time} (the final result is independent of the
specific choice), the diagram contributes the following integral:
\begin{equation}
I(\tau,{\bf k}) =\int\frac{d^3p}{(2\pi)^3}D_0(\tau,{\bf
p})D_0(\tau,{\bf k}-{\bf p}).
\end{equation}
The Fourier transform of $I(\tau,{\bf k})$ is obtained through the formula:
\begin{equation}
I(i\omega_e,{\bf k})=\int_0^{\beta}d\tau\,e^{-i\omega_e\tau}I(\tau,{\bf k}),
\end{equation}
where $\omega_e$ is  an external
Matsubara frequency.
By using the explicit form of the propagator given in Eq.(\ref{taupropag}),
we obtain (with ${\bf q}={\bf k}-{\bf p}$):
\begin{eqnarray}\label{oneloop1}
I(i\omega_e,{\bf k}) =
\int\frac{d^3p}{(2\pi)^3}\frac{1}{2\varepsilon_p}\frac{1}{2\varepsilon_q}
&\!&\!\!\!\!\!\left\{(1+n_{\varepsilon_p}+n_{\varepsilon_{q}})
\left(\frac{1}{i\omega_e+\varepsilon_p+\varepsilon_{q}}-\frac{1}{i\omega_e-\varepsilon_p-\varepsilon_{q}}\right)\right.\nonumber\\
&\,&+\left.
(n_{\varepsilon_p}-n_{\varepsilon_{q}})
\left(\frac{1}{i\omega_e-\varepsilon_p+\varepsilon_{q}}-\frac{1}{i\omega_e+\varepsilon_p-\varepsilon_{q}}\right)\right\}.
\end{eqnarray}
In order to arrive at this expression, we have used the fact that
$\omega_e$ is a
Matsubara frequency so that ${\rm e}^{i\beta\omega_e}=1$. We have
also used simple identities like
$1+n_{\varepsilon_k}=e^{\beta\varepsilon_k}n_{\varepsilon_k}$ in
order to eliminate the exponential factors resulting from the $\tau$
integration. Note that terms containing  products of statistical
factors, which appear in the initial stage of the
calculation, have cancelled out in the final formula.

The  expression  (\ref{oneloop1}) exhibits the contributions
of the various physical
processes that take place in the heat bath: pair creation or
annihilation, which occur also in the
vacuum, and scattering processes which take place only in the heat
bath.

The separation of  vacuum and  thermal contributions in Eq.(\ref{oneloop1})
is straightforward, and proceeds
in the same way as for the  fluctuation calculation above (see
Eq.~(\ref{I0I1})). The vacuum part is obtained by dropping in
Eq.~(\ref{oneloop1}) the terms which
contain a statistical factor:
\begin{equation}\label{PI0}
I^{(0)}(i\omega_e,{\bf k})=\int\frac{d^3p}{(2\pi)^3}
\frac{1}{2\varepsilon_p}\frac{1}{2\varepsilon_q}
\left(\frac{1}{i\omega_e+\varepsilon_p+\varepsilon_{q}}-\frac{1}{i\omega_e-\varepsilon_p-\varepsilon_{q}}\right).
\end{equation}
The thermal contribution, the sum of terms with one  statistical
factor, can be put in a compact
form  by exploiting the symmetries of the diagram in regrouping terms. One
easily gets:
\begin{eqnarray}\label{Ioneloop1}
I^{(1)}(i\omega_e,{\bf k})=2
\int\frac{d^3p}{(2\pi)^3}\frac{1}{2\varepsilon_p}n_{\varepsilon_{p}}
\left[D_0(i\omega_e-\varepsilon_p, {\bf k}-{\bf 
p})+D_0(i\omega_e+\varepsilon_p,{\bf k}-{\bf p})\right],
\end{eqnarray}
where we have used the expression (\ref{rho0D0}) of $D_0(\omega)$.

Note that in the calculations above it was essential to keep
$\omega_e$ as a Matsubara frequency
(so that e.g. ${\rm e}^{i\beta\omega_e}=1$). The final formula
however gives $I$ as an analytic
function which can be continued to all values of $\omega=i\omega_e$
in the complex plane, with
singularities on the real axis.

\subsubsection{Frequency representation}

\begin{figure}[htbp]
\begin{center}
\includegraphics[width=5cm]{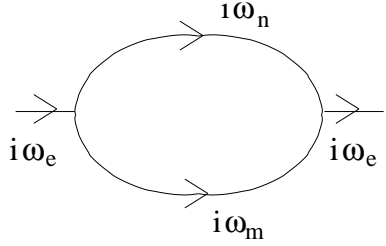}
\caption{One-loop self-energy in $\phi^3$ scalar field theory, frequency
representation.\label{fig:1loop_freq}}
\end{center}
\end{figure}

Alternatively, one may start from the energy
representation and label each line with independent Matsubara
frequencies $\omega_m$ and $\omega_n$ (see Fig.~\ref{fig:1loop_freq}).
Taking into account the energy conservation
at the vertices, we may set $\omega_m=\omega_e-\omega_n$, and obtain:
\begin{equation}\label{IoneloopMa}
I(i\omega_e,{\bf
k})=\frac{1}{\beta}\sum_n\int\frac{d^3p}{(2\pi)^3}D(i\omega_n,{\bf p})
D(i\omega_e-i\omega_n,{\bf k}-{\bf p}).
\end{equation}
Next, we use the spectral representation
(\ref{represspectrale}) for each propagator  and perfom the sum over
the Matsubara
frequency $\omega_n$. One gets (${\bf q}={\bf k}-{\bf p}$):
\begin{equation}\label{PIoneloop0}
I(i\omega_e,{\bf 
k})=\int\frac{d^3p}{(2\pi)^3}\int_{-\infty}^{\infty}\frac{dp_0}{2\pi}
\rho(p_0,{\bf
p})\int_{-\infty}^{\infty}\frac{dq_0}{2\pi}\rho(q_0,{\bf
q})\frac{n_{q_0}-n_{-p_0}}{p_0+q_0-i\omega_e}.
\end{equation}
Note that the numerator in Eq.~(\ref{PIoneloop0}) can be written also as
$n_{p_0}-n_{-q_0}$ ($=n_{q_0}-n_{-p_0}$).
At this point, we can use Eq.~(\ref{represspectrale}) to perform
trivially one of the energy integrals in
Eq.~(\ref{PIoneloop0}). One gets then:
\beq\label{IoneloopMa0}
I(i\omega_e,{\bf k})&=&
\int\frac{d^3p}{(2\pi)^3}\int_{-\infty}^{\infty}\frac{dp_0}{2\pi}\rho(p_0,{\bf
p})
[-n(-p_0)] D(i\omega_e-p_0,{\bf k}-{\bf p})\nonumber\\
&+&\int\frac{d^3p}{(2\pi)^3}\int_{-\infty}^{\infty}\frac{dq_0}{2\pi}\rho(q_0,{\bf 
k}-{\bf p})[n(q_0)]
D(i\omega_e-q_0,{\bf p}).
\eeq

The separation of the finite temperature contribution is now easily
achieved with the help of Eq.~(\ref{statfactor}). A simple calculation
allows us to recover Eq.~(\ref{PI0}) above for $I^{(0)}$. Note that $I^{(0)}$
can also be written as a 4-dimensional Euclidean integral:
\begin{equation}\label{euclideanoneloop}
I^{(0)}(i\omega_e,{\bf k})=\int\frac{d^4p}{(2\pi)^4}D(ip_0,{\bf
p})D(i\omega_e-ip_0,{\bf k}-{\bf p}).
\end{equation}
This is obvious from Eq.~(\ref{IoneloopMa}), but can be also verified
directly starting from Eq.~(\ref{IoneloopMa0}).  We shall often use 
the following notation for  integrals
such as that in Eq.~(\ref{euclideanoneloop}):
\begin{equation}\label{euclideanoneloop2}
I^{(0)}(K)=\int\frac{d^4P}{(2\pi)^4}D(P)D(K-P),
\end{equation}
with $K=(i\omega_e,{\bf k})$, $P=(ip_0,{\bf p})$.

   As for $I_1$, we shall write it as:
\begin{eqnarray}\label{PIoneloop1}
I^{(1)}(i\omega_e,{\bf k}) = 2 \int\frac{d^4p}{(2\pi)^4}\sigma(p_0,{\bf
p}) D(i\omega_e-p_0,{\bf k}-{\bf p}),
\end{eqnarray}
where $\sigma(p_0,{\bf p})$ is defined in
Eq.~(\ref{thermalrho}). A simple calculation, using the free spectral
function
allows us to recover Eq.~(\ref{Ioneloop1}).
We shall argue later that the formulae (\ref{IoneloopMa0}) and
(\ref{PIoneloop1})
can be written directly, i.e.,  without calculation, by using an
appropriate set of rules.

   In view of future developments, we redo now the sum over Matsubara
frequencies in a  way
that may look at this point artificially complicated. Let us keep the
labels $\omega_n$ and
$\omega_m$ of the two lines as they are in Fig.~\ref{fig:1loop_freq},
and write the resulting
product of denominators in the integral (\ref{IoneloopMa}) as:
\beq\label{id1l1}
\frac{1}{p_0-i\omega_n}\,\frac{1}{q_0-i\omega_m}=\frac{1}{p_0+q_0-i\omega_e}\left\{
\frac{1}{p_0-i\omega_n}+\frac{1}{q_0-i\omega_m}\right\}.
\eeq
This equation is valid if $\omega_n+\omega_m=\omega_e$ (which implies
in particular that
$\omega_e$ is a Matsubara frequency at this stage of the
calculation). But on the right hand side
the sum over the Matsubara frequencies can be done independently on
$\omega_n$ and $\omega_m$.
There is however one subtlety related to the fact that these
individual sums are ill-defined.
Following Gaudin \cite{Gaudin65}, we introduce a regulator and
rewrite Eq.~(\ref{id1l1}) as
\beq\label{id1l2}
\frac{{\rm e}^{i\omega_n\tau_n}}{p_0-i\omega_n}\frac{{\rm
e}^{i\omega_m\tau_m}}{q_0-i\omega_m}
=\frac{1}{p_0+q_0-i\omega_e}\left\{
\frac{{\rm e}^{i\omega_e \tau_m }\,{\rm 
e}^{i\omega_n(\tau_n-\tau_m)}}{p_0-i\omega_n}+\frac{{\rm 
e}^{i\omega_e\tau_n}\,{\rm
e}^{i\omega_m(\tau_m-\tau_n)}}{q_0-i\omega_m}\right\},
\eeq
where in the right hand side we have used the relation 
$\omega_n+\omega_m=\omega_e$
to express
each term as a function of the Matsubara frequency which sits in the
denominator. The sum over
Matsubara frequencies are now well defined: in the left hand side, it
is so even in the absence of
the regulator, and we can let $\tau_n$ and $\tau_m$ go to zero; in
the right hand side, the
sums are well defined provided we keep $\tau_n-\tau_m\ne 0$ when we
take the limit. Using the
formula (\ref{regularsum}) to perform the sums, we then easily
recover Eq.~(\ref{PIoneloop0}) with the two
ways of writing the  numerator corresponding to the two possible
limits $\tau_n-\tau_m\to 0"\pm$. This
method will be generalized in our next example.

\subsection{Two loop in one way}\label{sec:twoloop}

\begin{figure}[htbp]
\begin{center}
\includegraphics[width=5cm]{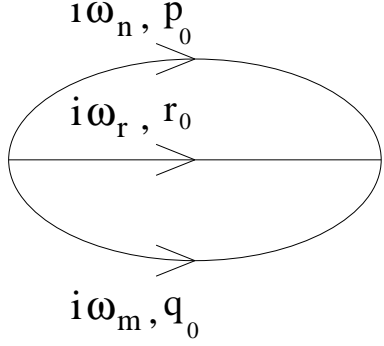}
\caption{A two-loop contribution to the pressure in scalar $\phi^3$
theory. To each line are
attached two labels: a Matsubara frequency and a real variable
representing the energy variable of the
spectral function of the propagator.
\label{fig:2looplabels}}
\end{center}
\end{figure}

Our next example is the 2-loop contribution to the free energy in
$\phi^3$ scalar field theory displayed
in  Fig.~\ref{fig:2looplabels}.  While a
   calculation through integrations in the imaginary time could be done
as easily as in the
previous case (there is again only one integration to be done), we go
here  directly to the
frequency representation and   proceed by labelling the various
internal lines of the
diagram with independent Matsubara frequencies.

First, we take into account the  conservation of
energy in order to work with independent frequencies, i.e., we set
$\omega_r=-\omega_n-\omega_m$. One is
then led to calculate  the  following sum-integral (we set
${\bf r}={\bf k}-{\bf p}-{\bf q}$):
\begin{equation}\label{Matsum20}
I=\frac{1}{\beta^2}\sum_n\sum_m\int\frac{d^3p}{(2\pi)^3}\int\frac{d^3q}{(2\pi)^3}D(i\omega_n,{\bf p})D(i\omega_m,{\bf q})D(-i\omega_n-i\omega_m,{\bf r}).
\end{equation}
Using as before the spectral representation for each
propagator, one ends up with the following sum
over  Matsubara frequencies:
\begin{eqnarray}\label{Matsum2}
\frac{1}{\beta^2}\sum_{n,m}\frac{1}{(p_0-i\omega_n)(q_0-i\omega_m)(r_0+i\omega_n+i\omega_m)}
=\frac{\left(-n_{q_0}+n_{-r_0}\right)\left(-n_{p_0}+n_{-q_0-r_0}\right)}{r_0+p_0+q_0}.
\end{eqnarray}
In order to obtain the right hand side,  we have performed the sum
over $\omega_m$ first and then that over
$\omega_n$. The statistical factor  $n_{-q_0-r_0}$ involving  a
linear combination  of frequencies
can  be transformed using the identity:
\begin{equation}
\left(-n_{q_0}+n_{-r_0}\right)
n_{-q_0-r_0}=\left(1+n_{-q_0}+n_{-r_0}\right)
n_{-q_0-r_0}=n_{-q_0}n_{-r_0}.
\end{equation}
One then obtains:
\begin{equation}\label{decomp1}
\frac{1}{\beta^2}\sum_{n,m}\frac{1}{(p_0-i\omega_n)(q_0-i\omega_m)(r_0+i\omega_n+i\omega_m)}
=\frac{n_{p_0}n_{q_0}-n_{p_0}n_{-r_0}+n_{-q_0}n_{-r_0}}{r_0+p_0+q_0}.
\end{equation}

There is a more systematic way to arrive directly at the expression
(\ref{decomp1}) in
which each statistical factor is function  of  the energy variable
carried by a single line (rather than involving sums of energy
variables as in Eq.~(\ref{Matsum2})). This requires leaving open the
choice of the independent
Matsubara frequencies so as to allow the calculation to proceed in as
a symmetrical way as
possible.  Let us then attach to the three  internal lines the
variables $\{p_0, q_0, r_0\}$ and
$\{\omega_n,
\omega_m,\omega_r\}$, where
$p_0, q_0, r_0$ are independent energy variables (the arguments of
the spectral functions) and $\omega_n, \omega_m,\omega_r$ are
Matsubara
frequencies constrained by the relation
\begin{equation}\label{contrainteomega}
\omega_n+\omega_m+\omega_r=0.
\end{equation}
There are three ways of parametrizing the solutions of this equation
(by choosing 2 of the 3 internal Matsubara frequencies as the independent
variables). Correspondingly, the fraction on the left hand side of
Eq.~(\ref{decomp1}) can be decomposed
in the following sum of three simpler fractions, each term involving
one choice of independent variables:
\begin{eqnarray}\label{treetwoloop}
\frac{1}{(p_0-i\omega_n)(q_0-i\omega_m)(r_0-i\omega_r)} & =
&\frac{1}{p_0+q_0+r_0}\nonumber\\
& & \hspace{-3cm} \times\left\{\frac{1}{(p_0-i\omega_n)(q_0-i\omega_m)}
+\frac{1}{(p_0-i\omega_n)(q_0-i\omega_r)}+\frac{1}{(p_0-i\omega_m)(q_0-i\omega_r)}\right\},\nonumber\\
\end{eqnarray}
the equality being valid when $\omega_n,\omega_m,\omega_r$ satisfy
the relation (\ref{contrainteomega}).
This decomposition (analogous to that in Eq.~(\ref{id1l1})) has a
diagrammatic interpretation
that we shall discuss more generally later.

This formula allows us to perform the sum over the
Matsubara frequencies in Eq.~(\ref{decomp1}) by summing in each term
of the right hand
side of (\ref{treetwoloop}) over the appropriate independent
variables.  As was the case in the
previous example (see Eq.~(\ref{id1l1})),  while the sum on the left
hand side is well defined
(since each independent frequency occurs there at least twice; see
e.g. (\ref{Matsum20})), this is
not so  in the right
    hand side. As we did in Eq.~(\ref{euclideanoneloop}),  we then
introduce a regulator in the form of
exponential factors  attached  to each line, and write:
\begin{eqnarray}\label{treetwoloop2}
\frac{e^{i\omega_n\tau_n}e^{i\omega_m\tau_m}e^{i\omega_r\tau_r}}
{(p_0-i\omega_n)(q_0-i\omega_m)(r_0-i\omega_r)} & =
&\frac{1}{p_0+q_0+r_0}\nonumber\\
& & \hspace{-5cm} \times\left\{\frac{e^{i\omega_n(\tau_n-\tau_r)}
e^{i\omega_m(\tau_m-\tau_r)}}{(p_0-i\omega_n)(q_0-i\omega_m)}+
\frac{e^{i\omega_n(\tau_n-\tau_m)}e^{i\omega_r(\tau_r-\tau_m)}}{(p_0-i\omega_n)(q_0-i\omega_r)}
+\frac{e^{i\omega_m(\tau_m-\tau_n)}e^{i\omega_r(\tau_r-\tau_n)}}{(p_0-i\omega_m)(q_0-i\omega_r)}\right\},\nonumber\\
\end{eqnarray}
where
$\tau_n$,
$\tau_m$,
$\tau_r$ are arbitrary times. In each term of the right hand side of
Eq.~(\ref{treetwoloop}),
   we have used Eq.~(\ref{contrainteomega}) to express the exponential
factors  in terms of the
relevant independent Matsubara frequencies.  As long as the various
combinations of time do not
vanish, the sums  over Matsubara frequencies are now well defined.
Once they are performed, we may
take the limit
$\tau\to 0$. The left hand side is  well defined when
$\tau\rightarrow 0$. In the right hand side, each term has a limit
    that depends on the way we take the limit  $\tau\rightarrow 0$ (see
Eq.~(\ref{regularsum})). Of
course, the sum of these 3 terms is independent of the way we take the
limit. Let us consider for example the limit:
\beq
\tau_n = 3\theta\qquad
\tau_m = 2\theta\qquad
\tau_r = \theta\qquad
\theta & \rightarrow & 0^{+}\, .
\eeq
We obtain then:
\begin{equation}\label{decomp3}
\frac{1}{\beta^2}\sum_{\{n,m,r\}}\frac{1}{(p_0-i\omega_n)(q_0-i\omega_m)(r_0-i\omega_r)}= 
\frac{n_{p_0}n_{q_0}-n_{p_0}n_{-r_0}+n_{-q_0}n_{-r_0}}{r_0+p_0+q_0},
\end{equation}
where the notation $\sum_{\{n,m,r\}}$ is meant to indicate that the
summation over the
Matsubara frequencies is constrained by Eq.~(\ref{contrainteomega}).
Eq.~(\ref{decomp3}) is
identical to Eq.~(\ref{decomp1}). Other choices of the limit would
lead to distinct but equivalent
expressions. This non uniqueness is of the same nature as that
discussed after Eq.~(\ref{PIoneloop0}).

We now return to the sum-integral (\ref{Matsum20}) and use
Eq.~(\ref{decomp1}) to write:
\begin{equation}\label{decomp5}
I=\int_p\int_q\int_{-\infty}^{\infty}\frac{dp_0}{2\pi}\frac{dq_0}{2\pi}
\frac{dr_0}{2\pi}\rho(p_0,{\bf p})\rho(q_0,{\bf q})\rho(r_0,{\bf
r})\frac{n_{p_0}n_{q_0}-n_{p_0}n_{-r_0}+n_{-q_0}n_{-r_0}}{r_0+p_0+q_0}.
\end{equation}
Note that the expression (\ref{decomp5}) is well defined even in
cases where the denominator
vanishes, i.e., when $r_0+p_0+q_0=0$. This is because the numerator
also vanishes (linearly) in this case, as
is evident from Eq.~(\ref{Matsum2}). However, we shall find useful to
be able to treat
separately the various terms occuring in the numerator. In order to
manipulate well defined quantities,
it is necessary to introduce a regularisation, such as for instance a
principal value prescription, or
adding a small imaginary part to the denominator, i.e., replacing  in 
Eq.~(\ref{decomp1})
$1/(r_0+p_0+q_0)$ by
$1/(r_0+p_0+q_0+i\alpha)$, where $\alpha$ is infinitesimal.  Of
course, it is only the sum of the three terms in Eq.~(\ref{decomp1}) 
that is independent of $\alpha$: individual contributions will contain
imaginary parts  depending on
$\alpha$, and will be different if we use instead a principal value
prescription. We adopt in the
following the regularisation which consists in adding a small
imaginary part to the denominator. Then
we can  use the expression (\ref{represspectrale}) of the propagator
in terms  of the spectral
function in order to perform some energy integrations, and rewrite
Eq.~(\ref{decomp5}) as the following
sum of three terms:
\beq\label{decomp4}
I&=&\int_p\int_q\left\{\int_{-\infty}^{\infty}\frac{dp_0}{2\pi}\frac{dq_0}{2\pi}
\rho(p_0,{\bf p})\rho(q_0,{\bf q}) n_{p_0} n_{q_0}
D(-p_0-q_0-i\alpha,{\bf r})\right.\nonumber\\
&\,&\qquad
+\int_{-\infty}^{\infty}\frac{dp_0}{2\pi}\frac{dr_0}{2\pi}\rho(p_0,{\bf
p}) \rho(r_0,{\bf r})
(-n_{p_0}) n_{-r_0} D(-p_0-r_0-i\alpha,{\bf r})\nonumber\\
&\,&\qquad
+\left.\int_{-\infty}^{\infty}\frac{dq_0}{2\pi}\frac{dr_0}{2\pi}\rho(q_0,{\bf
q})\rho(r_0,{\bf r})
n_{-q_0} n_{-r_0} D(-q_0-r_0-i\alpha,{\bf r})\right\}.
\eeq
   This expression can be written
directly by using the rules derived in the next section.

We turn now to our main goal which is to isolate the vacuum
contributions. To do so, we
express $I$ in terms of statistical factors with positive arguments.
To this aim, we start from
Eq.~(\ref{decomp5}), split  each statistical  factor in two pieces according to
Eq.~(\ref{statfactor}),
    and gather terms containing  respectively zero, one and two statistical
factors, that we denote respectively by $I^{(0)}$, $I^{(1)}$, and 
$I^{(2)}$.  We get:
\beq\label{intJ}
I & = &\int_p\int_q
\int_{-\infty}^{\infty}\frac{dp_0}{2\pi}\frac{dq_0}{2\pi}
\frac{dr_0}{2\pi}\rho(p_0,{\bf p})\rho(q_0,{\bf p})\rho(r_0,{\bf
r})\frac{1}{r_0+p_0+q_0+i\alpha}\qquad\qquad\nonumber\\
& &
\qquad\qquad\qquad\times\left\{\theta(-p_0)\theta(-q_0)-\theta(-p_0)\theta(r_0)+\theta(q_0)\theta(r_0)\right.\nonumber\\
& &\qquad
\qquad\qquad\left.+3\varepsilon(p_0)n_{|p_0|}\left[-\theta(-q_0)+\theta(r_0)\right]
+3\varepsilon(p_0)n_{|p_0|}\varepsilon(q_0)n_{|q_0|}\right\}.
\eeq

The term which contains no thermal factors is the vacuum contribution $I^{(0)}$
to $I$. (The terminology refers here  to the explicit temperature dependence;
  if the spectral density $\rho$ is not the free spectral density 
$\rho_0$, it may depend on the temperature, generating
  implicit temperature dependence in $I^{(0)}$.) Note that there is no
singularity in this term, the combination of $\theta$ functions in the
numerator vanishing when
$r_0+p_0+q_0=0$, as one can easily verify. Thus the $i\alpha$ may be
omitted. In fact, this vacuum
contribution may also be written  as an Euclidean integral,
which may be more convenient for its explicit calculation (at least
in the  case where $D=D_0$):
\begin{equation}
I^{(0)}=\int\frac{d^4p}{(2\pi)^4}\int\frac{d^4q}{(2\pi)^4}D(ip_0,{\bf
p})D(iq_0,{\bf q})D(-ip_0-iq_0,{\bf k}-{\bf p}-{\bf q}).
\end{equation}

\begin{figure}[htbp]
\begin{center}
\includegraphics[width=5cm]{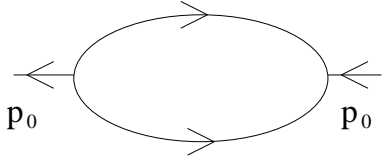}
\caption{A one-loop  subdiagram of the two-loop diagram of
Fig.~\ref{fig:2looplabels}  \label{fig:oneloop0}}
\end{center}
\end{figure}

In the thermal contributions, singularities may arise in individual
contributions and the $i\alpha$ must
be kept in the denominator. The term with one thermal factor contains
a vacuum one-loop contribution
which is represented in Fig.~\ref{fig:oneloop0}.
More precisely, we write:
\begin{equation}
\int_q\int_{-\infty}^{\infty}\frac{dq_0}{2\pi}\frac{dr_0}{2\pi}
\rho(q_0,{\bf q})\rho(r_0,{\bf
r})\,\frac{-\theta(-q_0)+\theta(r_0)}{p_0+q_0+r_0+i\alpha}=J_0(-p_0-i\alpha,{\bf 
p}),
\end{equation}
where $J_0$ is the one-loop integral given by
Eq.~(\ref{euclideanoneloop}) above (after proper analytic
continuation).
Then the contribution with one thermal factor is of the form:
\begin{equation}
I^{(1)}=3\int\frac{d^3p}{(2\pi)^3}\int_{-\infty}^{\infty}\frac{dp_0}{2\pi}\sigma(p_0,{\bf p})
J_0(-p_0-i\alpha,{\bf p}).
\end{equation}
The term with two thermal factors is given by:
\begin{equation}
I^{(2)}=3\int_p\int_q\int_{-\infty}^{\infty}\frac{dp_0}{2\pi}\frac{dq_0}{2\pi}\sigma(p_0,{\bf p})\sigma(q_0,{\bf p})D(-p_0-q_0-i\alpha,{\bf r}),
\end{equation}
and one can verify that the dependence on $\alpha$  disappears
in the sum $I^{(1)}+I^{(2)}$, as it should.

Summarizing, one can write the integral $I$ in Eq.~(\ref{intJ}) as
$I=I^{(0)}+I^{(1)}(\alpha)+I^{(2)}(\alpha)$, where the two temperature
dependent terms depend explicitly on the regulator $\alpha$, while
their sum does not. We should note also that we have been able to
write the subdiagrams involved in the calculation of $I^{(1)}$ and $I^{(2)}$
as simple analytical continuation of vacuum $n$-point functions,
$J_0(-p_0-i\alpha,{\bf p})$ and $D(-p_0-q_0-i\alpha,{\bf r})$ respectively.
In the rest of this paper, we shall examine under which conditions
such a
strategy can be generalized. But before we do that, it is useful to
recall the generalization of the method that we have
used to calculate the sums over the Matsubara frequencies.


\section{General rules \label{rules}}

\begin{figure}[htbp]
\begin{center}
\includegraphics[width=5cm]{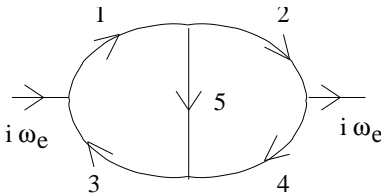}
\caption{A 2-loop contribution to the self-energy in $\phi^3$ scalar
field theory\label{fig1}}
\end{center}
\end{figure}

The rules that we are about to describe have been derived long ago by
M. Gaudin \cite{Gaudin65}, but his
   work seems to have been largely ignored
in the recent literature. We therefore find it appropriate to
recall here the  main steps involved in their derivation, without 
however going into all the subtleties of the
complete proof  which can be found in \cite{Gaudin65}. As we
proceed,
   in order to make the discussion more concrete, we shall carry along
a specific non trivial example,
that of the two-loop diagram of Fig.~\ref{fig1}.

Consider a general diagram $\Gamma$ with  $N_I$ internal lines, $N_V$
vertices and $N_L$ loops. In order to avoid complications with 
multiple poles, we assume $\Gamma$ to be
two-particle irreducible, i.e., one cannot isolate self-energy 
insertions on the internal lines of
$\Gamma$. We assume that each line of the diagram has been oriented 
and labelled (see the example in
Fig.~\ref{fig1}). The external frequencies will be denoted collectively by
$\{\omega_e\}$, and
the internal frequencies by
$\{\omega_i\}$ with $i=1,\cdots,N_I$. The evaluation of the diagram
involves that of the following sum over the Matsubara frequencies:
\begin{equation}
I\left\{i\omega_e\right\}=\frac{1}{\beta^{N_L}}\sum_{\left\{n_i\right\}}\prod_{i=1}^{N_I} D(i\omega_i),
\end{equation}
where the notation
$\sum_{\left\{n_j\right\}}$ stands for a sum restricted to  $N_L$
independent Matsubara frequencies (we leave aside the momentum 
integrals which play no role in the present
discussion). The  choice of independent  variables is at this point
left unspecified. The
$N_I$ internal Matsubara frequencies
$\omega_i$   obey
$N_V-1$ independent linear relations (also involving
$\left\{i\omega_e\right\}$):
\begin{equation}\label{constraints}
R_v\left\{i\omega_e,i\omega_i\right\}=0\mbox{, for $v=1,\dots,N_V-1$},
\end{equation}
which reflect the conservation of energies at each vertex. In $R_v$, a
frequency $\omega$ appears with a positive sign if it is attached
to a line which enters the vertex $v$, with a negative sign if the
corresponding line  leaves the vertex.
    The number of independent variables is  $N_I-N_V+1=N_L$.

We then express  each propagator in terms of its spectral function (see
Eq.~(\ref{represspectrale})), and  write:
\begin{equation}\label{I1}
I\left\{i\omega_e\right\}=\frac{1}{\beta^{N_L}}\sum_{\left\{n_i\right\}}
\prod_{i=1}^{N_I}\int\frac{dp_{i}^0}{2\pi}\rho(p^0_{i})\,\frac{1}{p^0_{i}-i\omega_i}.
\end{equation}
At this point, to each internal line $i$ of the diagram are attached
two energy variables: a real variable $p^0_i$, argument of a spectral
function, and a Matsubara frequency $\omega_i$.  While the real
variables $p^0_i$ are independent, this is not so for the Matsubara
frequencies which satisfy Eqs.~(\ref{constraints}). The problem is then to
compute the following sum:
\begin{equation}\label{ratfrac}
\frac{1}{\beta^{N_L}}\sum_{\left\{n_i\right\}}\prod_{i=1}^{N_I}
\frac{1}{p^0_{i}-i\omega_i},
\end{equation}
where the sum $\sum_{\left\{n_i\right\}}$ is restricted to Matsubara
frequencies satisfying Eqs.~(\ref{constraints}).

We proceed by generalizing Eq.~(\ref{treetwoloop}), and recall how
the system of linear equations
(\ref{constraints}) can be solved by exploiting the notion of tree
diagrams \cite{Gaudin65}.
Given a connected diagram $\Gamma$, a tree is a set of lines of
$\Gamma$ joining all the vertices
and   making a connected graph without loops.  It can be shown that
   each set of independent variables that can be chosen in order to
express the solutions of
Eqs.~(\ref{constraints}) can be associated with one of the trees that
can be identified on the
diagram  considered. We denote by
${\cal T}$ the set of lines which belong to a given tree  and by
$\bar{\cal T}$ the set of lines
which do not  belong to ${\cal T}$. There are $N_V-1$ lines in ${\cal
T}$ and  $N_L$
lines in  $\bar{\cal T}$. For a given tree ${\cal T}$, the $N_L$  independent
variables are the Matsubara frequencies attached to the lines of
$\bar{\cal T}$; we shall denote them by $\left\{\omega_{ l}\right\}$.
The remaining variables $\left\{\omega_{j}\right\}$, with
$j\in {\cal T}$, are linear combinations $\Omega_j$ of the
independent internal Matsubara frequencies
$\{\omega_{l}\}$  and  the external Matsubara frequencies
$\left\{\omega_e\right\}$:
\begin{equation}\label{treeformula}
j\in {\cal T},\,\,\,  l \in \bar{\cal T}, \qquad \omega_j=\Omega_j
\left\{i\omega_e,i\omega_{ l}\right\} .
\end{equation}
There is a simple way to read the values of the frequencies
$\Omega_j\{i\omega_e,i\omega_{l}\}$ on a graph: $
i\Omega_j$ is the algebraic sum of all the energies ($i\omega_{l}$
for the lines of $\bar {\cal T}$
and
$i\omega_e$ for the external lines)
   which flow through the (oriented) branch $j$ of ${\cal T}$.

We then have  the following formula that allows us to reduce the
rational fraction of
Eq.~(\ref{ratfrac}), when the $\omega_i$'s
satisfy Eqs.~(\ref{constraints}):
\begin{equation}\label{reduction}
\prod_{i=1}^{N_I}\frac{1}{p^0_{i}-i\omega_i}=\sum_{{\cal T }}\prod_{j\in
T}\frac{1}{p_j^0-i\Omega_j\left\{i\omega_e,p^0_{ l}\right\}}\prod_{
l\in \bar{\cal T}}\frac{1}{p^0_{l}-i\omega_{l}},
\end{equation}
where the sum over the trees corresponds to the sum over all possible
sets of independent internal Matsubara frequencies.
In this formula, the entries $i\omega_l$ in
$i\Omega_j\{i\omega_e,i\omega_l\}$ of Eq.~(\ref{treeformula}) have
been replaced by the
corresponding real energies $p_{l}^0$.

\begin{figure}[htbp]
\begin{center}
\includegraphics[width=14cm]{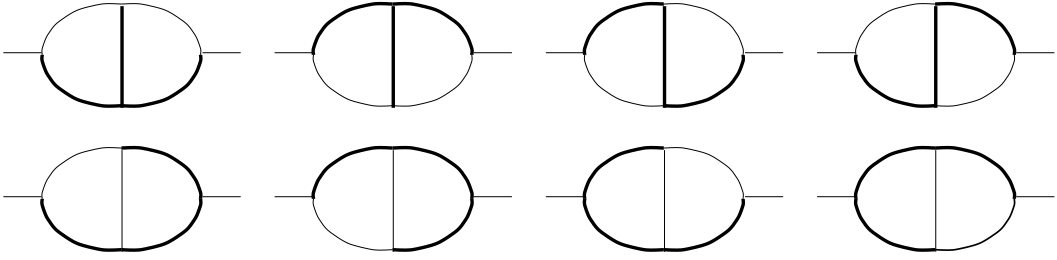}
\caption{The various trees of the two-loop diagram of
Fig. \ref{fig1} are represented by
thick lines. The thin lines are those of $\bar{\cal T}$: they carry
the independent Matsubara frequencies.\label{fig2}}
\end{center}
\end{figure}

As an illustration we show in Fig.~\ref{fig2} the various trees
corresponding to the 2-loop diagram in Fig.~\ref{fig1}. With the labelling
of
Fig.~\ref{fig1},  the
independent frequencies  for the first tree in Fig.~\ref{fig2} are
$\omega_1$ and $\omega_2$
and we have:
$i\Omega_3=p^0_1-i\omega_e$, $i\Omega_4=p^0_2-i\omega_e$,
$i\Omega_5=p^0_1-p^0_2$. Thus the  formula (\ref{reduction}) for the
first
tree yields:
\beq\label{denomfirsttree}
\prod_{i=1}^{5}\frac{1}{p^0_{i}-i\omega_i}\longrightarrow
\frac{1}{p^0_3-p^0_1+i\omega_e}\,\,
\frac{1}{p^0_4-p^0_2+i\omega_e}\,\,
\frac{1}{p^0_5-p^0_1+p^0_2}\,\,
\frac{1}{p^0_1-i\omega_1}\,\,
\frac{1}{p^0_2-i\omega_2}.
\eeq

At this point, the sum over the Matsubara frequencies,
Eq.ó(\ref{ratfrac}),  reduces, for
each tree, to:
\beq
\frac{1}{\beta^{N_L}}\prod_{l=1}^{N_L} \sum_{n_l}\frac{1}{p^0_{l}-i\omega_l},
\eeq
where now the $\omega_l$'s  are independent variables, and we have 
set $\omega_l= 2\pi n_l T$. These sums are
ill-defined in the absence of  a regulator. We proceed as in the
   examples of the previous section and attach to each internal line a
time $\tau_i$ that
we
shall let go to zero at the end of the calculation, transforming
Eq.~(\ref{reduction}) into:
\begin{equation}\label{reduction2}
\prod_{i=1}^{N_I}\frac{{\rm
e}^{i\omega_i\tau_i}}{p^0_{i}-i\omega_i}=\sum_{{\cal T }}\left({\rm
e}^{i\omega_eT_e}\prod_{j\in
{\cal T }}\frac{1}{p_j^0-i\Omega_j\left\{i\omega_e,p^0_{ l}\right\}}\prod_{
l\in \bar{\cal T}}\frac{{\rm
e}^{i\omega_l T_l}}{p^0_{l}-i\omega_{l}}\right),
\end{equation}
where we have set:
\begin{equation}
\sum_i \omega_i\tau_i=\sum_{l\in\bar{T}}\omega_lT_l+\omega_e T_e.
\end{equation}
Both $T_e$ and $T_l$ are linear combinations of the times $\tau_i$.
We shall need in fact only  $T_l$, and this is determined by the
following simple rule.

First we observe that each line $l$ of $\bar{\cal T}$ defines a
unique loop of $\Gamma$. To
determine the linear combination
$T_l$ we consider successively all the lines of the loop
$l$:  a line
$k$ in this loop contributes $+\tau_k$ if it is oriented as the line $l$ and
$-\tau_k$ in the opposite case. Thus, in the example
discussed above, there are two loops. The phase factor reads
$\sum_i\omega_i\tau_i=\omega_1T_1+\omega_2 T_2+\omega_e T_e$ with
$T_1=\tau_1+\tau_3+\tau_5$, $T_2=\tau_2+\tau_4-\tau_5$ and
$T_e=-\tau_3-\tau_4$ (to obtain this result we have expressed
$\omega_3$,
$\omega_4$ and $\omega_5$ in terms of $\omega_1$, $\omega_2$ and
$\omega_e$, the independent
variables corresponding to the first tree in Fig.~\ref{fig2}).

    We are now ready to perform the sum over independent Matsubara
frequencies in Eq.~(\ref{reduction2}).
    The left hand side is well defined, even in the absence of a
regulator, so that the limit  $\tau_i\to 0$ can be taken, and
the result is independent of the ways the various $\tau_i$'s approach
0.  In the right hand
side, the regulator matters, and the results of individual sums
depend on the  sign $\epsilon_l$ of $T_l$ (see Eq.~(\ref{regularsum})):
\begin{equation}
\frac{1}{\beta}\sum_{n_l}\frac{e^{i\omega_{l}T_{l}}}{p^0_{l}-i\omega_{l}}
=\epsilon_l
n_{\epsilon_l p_l^0}\, e^{p^0_{l}T_{l}}.
\end{equation}
The factor $e^{i\omega_e T_e}$ has no
influence in the limit $\tau_i\rightarrow 0$ and we can forget it; 
the same remark applies to the factor $e^{p^0_{l}T_{l}}$ in the right 
hand side.
Thus, the sum-integral (\ref{I1}) can be written as follows:
\begin{equation}\label{pre_Gaudin}
I\left\{i\omega_e\right\}=\int\prod_{i=1}^{N_I}
\frac{dp^0_i}{(2\pi)}\rho(p_{i}^0)\sum_{{\cal T}}\left(\prod_{j\in
{\cal T}}\frac{1}{p^0_j-i\Omega_j}\prod_{l\in
\bar{\cal T}}\epsilon_l n_{\epsilon_l p_{l}^0}\right).
\end{equation}
This formula is essentially that derived by Gaudin \cite{Gaudin65}. 
It can be translated into  a set of rules
listed below.

{\noindent} {\bf Rules}

\begin{enumerate}

\item{  Determine the family  of all the trees $\cal T$ that can be
drawn on $\Gamma$. A tree is
a connected set of lines of $\Gamma$ which joins all the vertices and
contains no loop. A tree
contains $N_V-1$ lines, with $N_V$ the number of vertices of
$\Gamma$. We call $\bar{\cal T}$
the set of lines of $\Gamma$ which do not belong to ${\cal T}$. The
external lines play no role in
the determination of the trees: they  belong neither to ${\cal T}$,
nor to $\bar{\cal T}$. }

\item{ Specify the orientation of each line  of $\Gamma$, and affect
to each line $k$, and once
and for all,  a positive number
$\tau_k$. We  define the orientation of a loop by the following
rule. Given a tree ${\cal
T}$, consider the loop $l$ associated to the line $l$ of $\bar{\cal
T}$. The orientation of the
loop is the algebraic sum of the $\tau_k$ carried by each
line of the loop, with $\tau_k$
counted positively if the line
$k$ is oriented as the line $l$, and negatively (as $-\tau_k$) in
the opposite case. The choice of the
$\tau_k$ must be such that the orientation of each loop that can be
drawn on $\Gamma$ is non vanishing (it is always possible to do so). 
Note that loops and their  orientations may
change depending on the tree one begins with.  }
\item{  The contribution of the sum over Matsubara frequencies to
$\Gamma$ is then given by the
formula (\ref{pre_Gaudin}). It is the sum of the contributions of the
various trees ${\cal T}$. To
each line
$l$ of
$\bar{\cal T}$ is associated an integral over the energy $p^0_l$ with
the weight
$\rho(p_{l}^0)\epsilon_l n_{\epsilon_l p_l^0}$ where
$\epsilon_l=+ 1$ or $-1$ depending on whether the orientation of the
loop $l$ is, respectively, positive
or negative. To each line $j$ of ${\cal T}$ is associated a
factor  $\rho(p_j^0)\, (p_j^0-i\Omega_j)^{-1}$ where
$\Omega_j$ is determined as follows: given the orientation of the
line $j$, $i\Omega_j$ is the
algebraic sum of the quantities
$i\omega_e$ and $p^0_l$, carried respectively by the external lines
and the lines of $\bar{\cal
T}$, which flow through the line $j$ in the direction specified on the
line $j$. Note  that various choices of $\tau_k$ may lead  to 
seemingly different, but
equivalent, expressions (since the numerators depend on the 
orientation of the loops, and hence on the specific choice of the
$\tau_k$'s). We emphasize that the ambiguity concerns only the signs 
in front of the statistical factors and
their arguments. In particular, it is important to observe that, in 
all cases, each statistical factor
carries a single frequency attached to a line (and not sums of such 
frequencies).  }

\end{enumerate}

These rules apply indifferently at finite temperature and
at zero temperature. In the latter case,
   the factor to be
associated to each loop integral is $-\rho(p_{l}^0)\epsilon_l
\theta({-\epsilon_l
p_l^0})$ (rather than $\rho(p_{l}^0)\epsilon_l n_{\epsilon_l
p_l^0}$).

One can verify that these rules are satisfied on the
examples discussed in the
previous section (see Eqs.~(\ref{PIoneloop0}) and (\ref{decomp5})).
To apply them to our two-loop example, we first define a set
of regulators $\tau_k$; in the present case the choice
$\tau_k=k\theta^+$  (with $k$ an integer and $\theta\to 0$) is a 
possible one. Then we can compute for
instance  the numerator for the first tree in Fig.~\ref{fig2}. Consider
the loop involving the
line 1,  i.e., the set of lines $\left\{1,5,3\right\}$; its orientation,
$(1+5+3)\theta$, is positive, so that the corresponding contribution 
to the numerator
is $n_{p^0_1}$. The orientation of the loop $\left\{2,4,5\right\}$ is 
also positive ($(2+4-5)\theta$),
resulting  in the   contribution $n_{p^0_2}$ to the numerator. Combining  with
the denominator obtained from Eq.~(\ref{denomfirsttree}), we obtain
the contribution of the first tree as:
\begin{eqnarray}
\frac{n_{p^0_1}n_{p^0_2}}{(p^0_3-p^0_1+i\omega_e)(p^0_5+p^0_2-p^0_1)(p^0_4-p^0_2+i\omega_e)}.
\end{eqnarray}
Repeating this simple procedure for all the trees in Fig.~\ref{fig2} we obtain:
\begin{eqnarray}\label{eq:Gaudin_ex}
I(i\omega_e)  =
\int_{12345}& \,&
\left\{\,\,\frac{n_1n_2}{(p^0_3-p^0_1+i\omega_e)(p^0_5+p^0_2-p^0_1)(p^0_4-p^0_2+i\omega_e)}\right.\nonumber\\
& + &
\hspace{0.25cm}\frac{n_3n_4}{(p^0_1-p^0_3-i\omega_e)(p^0_5+p^0_4-p^0_3)(p^0_2-p^0_4-i\omega_e)}\nonumber\\
& + &
\hspace{0.25cm}\frac{n_3n_2}{(p^0_1-p^0_3-i\omega_e)(p^0_5+p^0_2-p^0_3-i\omega_e)(p^0_4-p^0_2+i\omega_e)}\nonumber\\
& + &
\hspace{0.25cm}\frac{n_1n_4}{(p^0_3-p^0_1+i\omega_e)(p^0_5-p^0_1+p^0_4+i\omega_e)(p^0_2-p^0_4-i\omega_e)}\nonumber\\
& + &
\hspace{0.25cm}\frac{n_1n_{-5}}{(p^0_3-p^0_1+i\omega_e)(p^0_2+p^0_5-p^0_1)(p^0_4+p^0_5-p^0_1+i\omega_e)}\nonumber\\
& + &
\hspace{0.25cm}\frac{n_3n_{-5}}{(p^0_1-p^0_3-i\omega_e)(p^0_4+p^0_5-p^0_3)(p^0_2-p^0_3+p^0_5-i\omega_e)}\nonumber\\
& + &
\hspace{0.25cm}\frac{n_2n_5}{(p^0_3-p^0_2-p^0_5+i\omega_e)(p^0_1-p^0_2-p^0_5)(p^0_4-p^0_2+i\omega_e)}\nonumber\\
& + &
\hspace{0.25cm}\left.\frac{n_4n_5}{(p^0_1-p^0_4-p^0_5-i\omega_e)(p^0_3-p^0_4-p^0_5)(p^0_2-p^0_4-i\omega_e)}\right\},\nonumber\\
\end{eqnarray}
with the short-hand notations $n_i=n_{p^0_i}$ and $n_{-i}=-n_{-p^0_i}$ and
\beq\label{shortnotation}
\int_{12345}\equiv\int\prod_{i=1}^5\frac{dp^0_i}{2\pi}\rho(p^0_i).
\eeq

\section{Expansion in the number of statistical factors}\label{section:thermal}

We now return to our main goal, which is to isolate vacuum 
contributions in a general Feynman diagram
at finite temperature. The rules of the previous section enable us to 
do that easily. They
also allow us to see that it is not always possible to identify the 
vaccum subdiagrams with analytic
continuation of simple vacuum amplitudes.

To proceed with the separation of explicit temperature dependent 
contributions, we use   Eq.~(\ref{statfactor})
to isolate  the temperature dependent factor in
$\epsilon_l n_{\epsilon_l  p^0_l}$:
\begin{equation}
\epsilon_l  n_{\epsilon_l  p^0_l}=-\epsilon_l  \theta(-\epsilon_l
p^0)+\epsilon(p^0)n_{|p^0|}.
\end{equation}
   Next, in the contribution of each tree in Eq.~(\ref{pre_Gaudin}), we
replace the quantity $\epsilon_l  n_{\epsilon_l  p^0}$ attached to 
each line $l$
of
$\bar{\cal T}$ by its expression above, and separate the various
terms thus obtained. One gets then, for each tree,
$2^{N_L}$ contributions  containing terms with $0,\dots,N_L$ factors
$\epsilon(p^0_l)n_{|p^0_l|}$. Diagrammatically the operation is illustrated in
Fig.~\ref{fig4} for  our two-loop example: the total number of
contributions is $2^2\times
8=32$, each tree generating  one vacuum contribution, two
contributions with one statistical
factor, and one contribution with two statistical factors. The lines
of $\bar{\cal
T}$ carrying vacuum factors $-\epsilon_l  \theta(-\epsilon_l  p^0)$
are represented by thin lines; we shall call them ``vacuum lines''. 
Those carrying a statistical factor
$\epsilon(p^0_l)n_{|p^0_l|}$ are represented by  dotted lines; we 
shall call them ``thermal lines''. Note that in
each column in Fig.~\ref{fig4}, a given tree occurs once and only once.

In order to proceed further, we need to analyze the contributions of 
subsets of terms, for instance those which
contain a given thermal line. In other words, as suggested by the way 
the various diagrams are grouped in
Fig.~\ref{fig4}, we  wish to give a meaning to sums of terms
which involve only a subset of the trees of the diagram, that is,  we 
would like to manipulate independently all
the trees.  This raises problems that we now discuss.

\subsection{Regularized summation over the tree diagrams}\label{section:reg}

   Returning to the formula (\ref{pre_Gaudin}), we note that some 
denominators vanish for some
  particular values of the variables $p^0_j$, leading to potential 
singularities. However such singularities
are fictitious: indeed,  the  sum of all tree contributions
   is well defined even when denominators vanish (for an example, see 
Sect. \ref{sec:twoloop}). That
there are, in the original diagram, no singularities  associated with vanishing
denominators is clear in the time representation:
   a vanishing denominator would correspond in fact to an integral
proportional to
$\beta$, leading to no denominator! (We should not confuse such 
fictitious singularities with those
which may occur for certain real values of the external frequencies 
and which are associated with physical
processes). Now, since denominators can vanish in individual trees 
for some values of the integration variables
$p_i^0$, it is not  possible to do the integration
   over the $p_i^0$ before doing the sum over trees.  If we wish  to do so
and be able
to manipulate independently the contributions of the various trees,
   we need to introduce a regularization. We have met this problem
after Eq.~(\ref{decomp5}),
and we shall proceed similarly in the general case by attaching a small
imaginary part to the various denominators. There is some
arbitrariness in doing that, the only  constraint being that the
final result
   (i.e. including the sum over all trees) should be independent of the
choice of regulators. This constraint  implies in particular  that
a given denominator  occuring in different trees must carry
everywhere the same imaginary part. One way to guarantee this is to
add a small
imaginary part $i\alpha_j$ to all the variables $p^0_j$ of the 
internal lines \cite{BaDeDo}. Note that the regulators thus 
introduced may not
be all needed (nor chosen independently). For instance, in the 
expression (\ref{eq:Gaudin_ex}) for the
two-loop example, there are only two ``dangerous'' denominators, 
namely $(p_5^0+p_2^0-p_1^0)$ and
$(p_5^0+p_4^0-p_3^0)$. The other denominators contain the imaginary 
frequency $i\omega_e$ and cannot vanish.
Thus we need a priori only two regulators, namely the combinations 
$\alpha_5+\alpha_2-\alpha_1$ and
$\alpha_5+\alpha_4-\alpha_3$. But these particular combinations 
should not vanish, which places a constraint
on the choice of the $\alpha_j$.

Assuming such a regularization, we can perform trivially, in
Eq.~(\ref{pre_Gaudin}), the $N_I-N_L$ integrals over the
spectral densities attached to the lines of a given tree so as to  reconstruct  a
propagator $D$ for each line of the tree. The final result reads:
\begin{equation}\label{Gaudin}
I\left\{i\omega_e\right\}=\sum_{{\cal T}}\int\prod_{l\in
\bar{{\cal T}}}\frac{dp^0_l}{(2\pi)}\rho(p_{l}^0)\epsilon_l
n_{\epsilon_l p_l^0}\prod_{j\in
{\cal T}}D (i\Omega_j;\alpha),
\end{equation}
where $\alpha$ denotes collectively the set of regulators.

\begin{figure}
\begin{center}
\includegraphics[width=10cm]{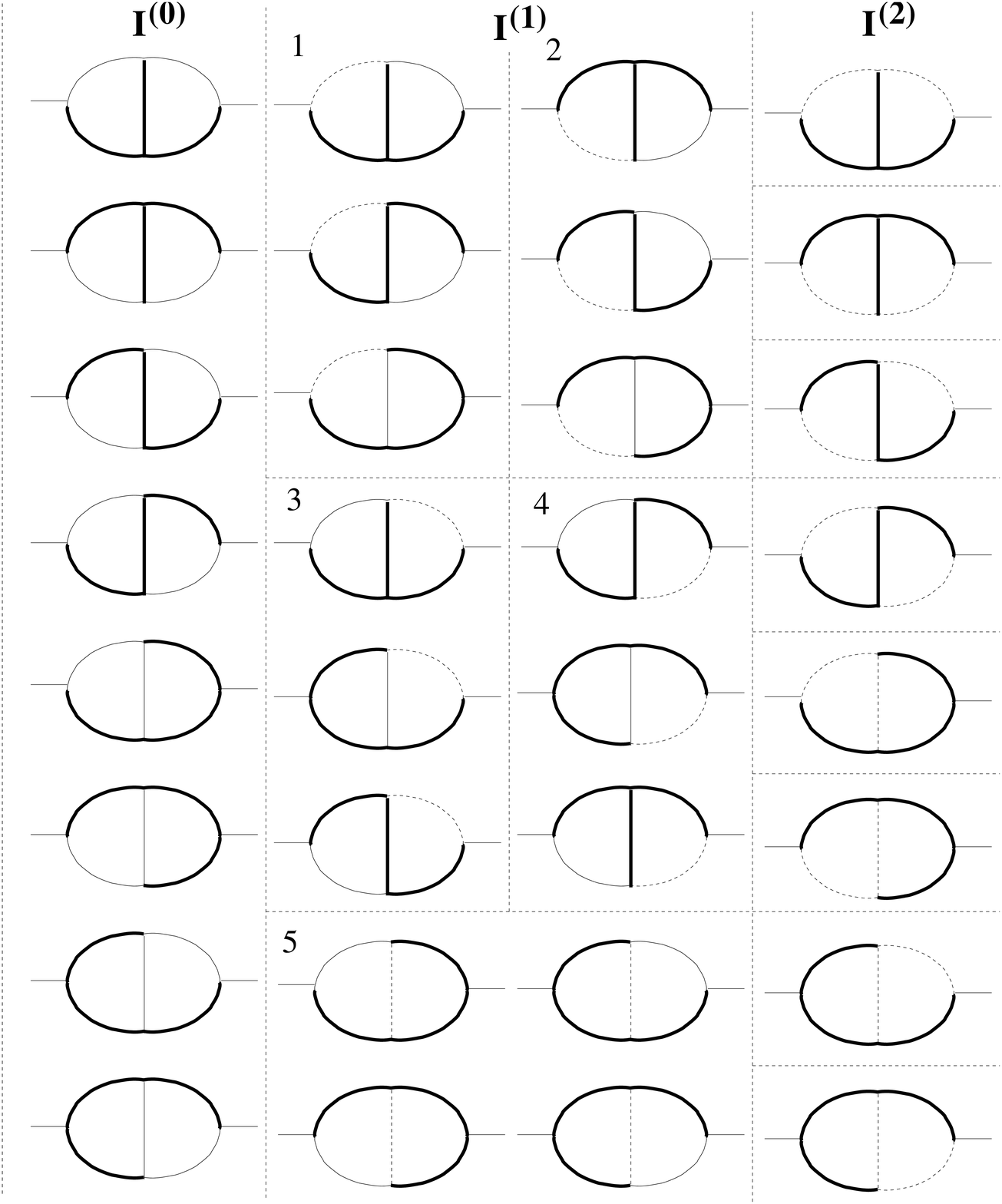}
\caption{The contributions to the 2-loop diagram of Fig.~\ref{fig1}
with zero ($I^{(0)}$), one ($I^{(1)}$) and two ($I^{(2)}$) thermal
factors. The thick lines represent the trees. The thin lines are
associated with zero temperature numerators, the dotted lines
with thermal factors. The first column contains all the zero
temperature contributions. The last column contains the
contributions with two thermal factors. The columns in the middle
contains the contributions with one thermal factor, grouped
by blocks where the thermal factor is attached to a given line of the
diagram.\label{fig4}}
\end{center}
\end{figure}

At this point we return to the general discussion and note that a
given tree generates $C_{N_L}^{N_l}$ terms with $N_l$ factors
$\sigma(p^0_l)\equiv \rho(p^0_l)\epsilon(p^0_l)n_{|p^0_l|}$ ($0\le 
N_l\le N_L$) . For a given $N_l$
we consider all the possible sets $\mathcal{A}$ of thermal lines 
(carrying $\sigma$ factors and  which we
label by the variables $p_a^0$). The  complementary sets 
$\bar{\mathcal{A}}$ are
$(N_I-N_l)$-loop connected subdiagrams (in general one
line-reducible). Examples are given in  Fig.~\ref{fig4} (for
instance, the various sets ${\cal A}$ with one statistical factor are
the five blocks labelled $1,2,\cdots,5$ in the middle columns of
Fig.~\ref{fig4}).
We then write:
\begin{eqnarray}\label{eq:decomp2}
I\left\{i\omega_e\right\} & = &
\sum_{N_l=0}^{N_L}\sum_{\mathcal{A}_{N_l}}
\int\prod_{a\in\mathcal{A}_{N_l}}\frac{dp^0_a}{2\pi}\sigma(p^0_a)I_{\mathcal{A}_{N_l}}
\left\{i\omega_e,p^0_a;\alpha\right\}\nonumber\\
& = & \sum_{N_l=0}^{N_L} \, I^{(N_l)}\{\omega_e;\alpha\}\, ,
\end{eqnarray}
where $I_{\mathcal{A}_{N_l}}\left\{i\omega_e,p^0_a;\alpha\right\}$ is the
sum of all the contributions to $ I\left\{i\omega_e\right\} $
which contain  the same set ${\cal A}_{N_l}$ of $N_l$ thermal lines, while
$I^{(N_l)}\left\{\omega_e;\alpha\right\}$ is the sum of all the 
contributions containing $N_l$ statistical
factors. In our  two-loop example, the sets
${\cal A}$ contain 0,1 or 2 thermal lines, and we can write, more explicitly:
\begin{eqnarray}\label{eq:decomp}
I\left\{i\omega_e\right\} & = & I^{(0)}\left\{i\omega_e;\alpha\right\}+I^{(1)}\left\{i\omega_e;\alpha\right\}+I^{(2)}\left\{i\omega_e;\alpha\right\},\eeq
with
\beq
  I^{(1)}\left\{i\omega_e;\alpha\right\} =
\sum_a\int\frac{dp^0_a}{2\pi}\sigma(p^0_a)I_a\left\{i\omega_e,p^0_a;\alpha\right\},
\eeq
and
\beq
I^{(2)}\left\{i\omega_e;\alpha\right\}=\sum_{(a,b)}\int\frac{dp^0_a}{2\pi}\frac{dp^0_b}{2\pi}
\sigma(p^0_a)\sigma(p^0_b)I_{ab}\left\{i\omega_e,p^0_a,p^0_b;\alpha\right\},
\end{eqnarray}
where the subscripts $a,b,\cdots$ label the selected thermal lines.
This formula  has a simple interpretation. The terms with one
statistical factor are obtained by replacing successively each internal
line by a thermal line: the five ways to do this correspond to the 
five blocks in the middle columns of
Fig.~\ref{fig4}. Similarly for the
terms with two statistical factors: the sum is  over all pairs 
$(a,b)$ such that the remaining lines constitute
a tree. Clearly these contributions are in one-to-one correspondence 
with the various trees.
  The dependence on $\alpha$  in $I_a$ and $I_{ab}$
reminds us that the separation of contributions with different
numbers of thermal factors is well defined except for specific values
of the energies ${p_a^0}$ for which
denominators vanish: at these points combinations of statistical
factors vanish, destroying the classification of the contributions 
according to the number of statistical factors they contain.

We will associate to a given $\mathcal{A}_{N_l}$ a $n$-point function (with
$n=2N_l+N_e$,
where $N_e$ is the number of external lines)
$J_{\mathcal{A}_{N_l}}\left\{i\omega_e,i\omega_a,i\omega'_a\right\}$
corresponding to the computation of the
diagram
$\bar{\mathcal{A}}_{N_l}$ in the imaginary time formalism, at zero
temperature. This diagram is obtained by cutting the lines of 
$\mathcal{A}_{N_l}$ and attributing to the two
ends of the cut line, carrying initially the real variable $p^0_a$, two independent 
complex variables
$i\omega_a$ and $i\omega_a'$. The question to be addressed in this 
section  is whether the regularized integral
$I_{\mathcal{A}_{N_l}}\left\{i\omega_e,p^0_a;\alpha\right\}$ can be 
considered as an analytic continuation of
$J_{\mathcal{A}_{N_l}}\left\{i\omega_e,i\omega_a,i\omega_a'\right\}$ 
for suitably chosen variables $i\omega_a$
and
$i\omega_a'$.

Let us first verify that $I_{A_{N_l}}$ and $J_{A_{N_l}}$ are given by (almost)
identical rules. Consider a particular set  $\mathcal{A}$ with $N_l$
thermal lines. There exists a tree ${\cal T}$ on $\Gamma$ such that
the thermal  lines belong to $\bar{\cal T}$, which also contains $N_L-N_l$
vacuum  lines to which are associated factors
$-\epsilon_j\theta(-\epsilon_j p^0_j)$. Consider the set ${\cal 
T}_{\cal A}$ of  all the trees ${\cal
T}'$ whose complements contain the same thermal lines. Clearly ${\cal 
T}_{\cal A}$ contains  all the trees
contributing to the $n-$point function
$J_{\mathcal{A}}\left\{i\omega_e,i\omega_a,i\omega_a'\right\}$. Thus the
denominators of $J_A$ have the same structure as those of $I_A$, they 
differ solely in that in
$J_{\mathcal{A}}\left\{i\omega_e,i\omega_a,i\omega'_a\right\}$ we have 
attached independent complex variables
on the external lines, while  in 
$I_{\mathcal{A}}\left\{i\omega_e,p^0_a;\alpha\right\}$
there is a unique real frequency $p^0_a$ attached to both ends of the 
thermal line (to within the $i\alpha_a$ inherited
form the regularization). As for the numerators they are identical, 
to within the usual ambiguity related to the
choice of the loop orientations. This follows from the fact that the 
$\sigma$ factors do not depend on the choice
of the orientation: thus the sign is determined by  the orientation 
of the vacuum lines only, and these are
the lines of $\bar{\cal A}$.

To proceed now it is best to look at
specific examples. We shall consider next our two-loop example for 
which we can carry through successfully our
analysis, and show indeed that the separation of vacuum contributions 
allows for a very simple discussion of
ultraviolet divergences. At the end of this section, we shall discuss 
a counter-example showing that it is not
always possible to do so.

\subsection{The two-loop example}

Consider then the various contributions to the two-loop diagram, as 
displayed in  Fig.~\ref{fig4}
The first column in Fig.~\ref{fig4} lists the terms with no thermal 
line. Their sum $I_0$ is nothing but
   the zero temperature limit of the two-loop diagram. It can be
calculated with the rule
given above. Alternatively, it can be written as an Euclidean
integral (i.e. not performing the frequency integral first, but doing
the
calculation with covariant techniques). One gets then:
\beq
I^{(0)}(K)=\int\frac{d^4P}{(2\pi)^4}\int\frac{d^4Q}{(2\pi)^4}D(P)D(P-K)D(P-Q)D(Q-K)D(Q),
\eeq
where the notation is that of Eq.~(\ref{euclideanoneloop2}).

Consider next the sum of the  contributions with one thermal line 
that are associated with
  the diagrams of  the bloc 1 in Fig. \ref{fig4}). We write this as
\beq
\int\frac{dp^0_1}{2\pi}\sigma(p^0_1)I_1(i\omega_e,p^0_1;\alpha),
\eeq
with
\beq\label{I1analytic}
I_1(i\omega_e,p^0_1;\alpha)
=D(p^0_1+i\alpha_1-i\omega_e)L(i\omega_e,p^0_1;\alpha).
\eeq
A diagrammatic representation of $I_1(i\omega_e,p^0_1;\alpha)$ is 
given in Fig.~\ref{fig2a} below.
In $I_1$, we have isolated the common propagator 
$D(p^0_1+i\alpha_1-i\omega_e)$, and $L$ is defined by
\begin{eqnarray}\label{eq:I1Gaudin}
L(i\omega_e,p^0_1;\alpha)\equiv
\int_{245}&  &\!\!\!\left\{
\frac{-\theta(-p^0_2)}{(p^0_5+p^0_2-p^0_1)(p^0_4-p^0_2+i\omega_e)}\right.\nonumber\\ 
& + &
\frac{-\theta(-p^0_4)}{(p^0_5-p^0_1+p^0_4+i\omega_e)(p^0_2-p^0_4-i\omega_e)}\nonumber\\ 
& + &\left.
\frac{\theta(p^0_5)}{(p^0_2+p^0_5-p^0_1)(p^0_4+p^0_5-p^0_1+i\omega_e)}\right\}.
\end{eqnarray}
The notation $\int_{245}$ is that introduced in 
Eq.~(\ref{shortnotation}). In the denominators, it is
understood that all the variables $p^0_j$ are shifted by a small
imaginary part ($p^0_j\to p^0_j+i\alpha_j $), in agreement with the 
regularization introduced in the previous subsection.

\begin{figure}
\begin{center}
\includegraphics[width=3cm]{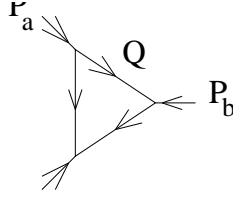}
\caption{One-loop contribution to the three-point function.\label{fig:4}}
\end{center}
\end{figure}

   At this point we
consider  a related diagram, that of the three-point function in 
Fig.~\ref{fig:4}, computed at zero
temperature, in imaginary time (we use  the convention of incoming 
external momenta):
\begin{equation}\label{Lambda0}
\Lambda(P_a,P_b)=\int\frac{d^4Q}{(2\pi)^4}D(P_a-Q)D(Q+P_b)D(Q).
\end{equation}
The notation for quadrimomenta is the same as in 
Eq.~(\ref{euclideanoneloop2}), that is,
$P_a\equiv(i\omega_a,{\bf p}_a)$,
$Q\equiv (iq_0,{\bf q})$.  The integral
in Eq.~(\ref{Lambda0}) can also be calculated by applying the rules 
of Sect.~\ref{rules} at zero temperature.
One gets (omiting the spatial momenta):
\begin{eqnarray}\label{eq:J1Gaudin}
\Lambda(i\omega_a,i\omega_b) = \int_{245}&  &\!\!\!\left\{
\frac{-\theta(-p^0_2)}{(p^0_5+p^0_2-i\omega_a)(p^0_4-p^0_2-i\omega_b)}\right.\nonumber\\ 
& + &
\frac{-\theta(-p^0_4)}{(p^0_5+p^0_4-i\omega_a-i\omega_b)(p^0_2-p^0_4+i\omega_b)}\nonumber\\ 
& + &\left.
\frac{\theta(p^0_5)}{(p^0_2+p^0_5-i\omega_a)(p^0_4+p^0_5-i\omega_a-i\omega_b)}\right\}.\nonumber\\ 
\end{eqnarray}
Thus defined, $\Lambda(i\omega_a,i\omega_b)$ is an analytic function 
of the variables $i\omega_a,i\omega_b$,
with singularities on the planes defined by ${\rm Im}
(i\omega_a)=0 $,  ${\rm Im} (i\omega_a)=0$ or ${\rm Im} 
(i\omega_a+i\omega_b)=0$. Alternatively, if one sets
$i\omega_a=p^0_a+i\alpha_a$, $i\omega_b=p^0_a+i\alpha_b$,
there are six domains of analyticity, depending on the relative signs
of $\alpha_a, \alpha_b$ and $\alpha_a+\alpha_b$. Now, it is not 
difficult to find a set of regulators making
it possible to identify  $L$ with $\Lambda$ in one of its domains of 
analyticity.  By comparing
Eqs.~(\ref{eq:J1Gaudin}) and (\ref{eq:I1Gaudin}),  one finds the relations:
\beq
i\omega_a=p_1^0+i(\alpha_1-\alpha_2-\alpha_5)\qquad
i\omega_b=-i\omega_e+i(\alpha_2-\alpha_4).
\eeq
By
choosing $\alpha_2-\alpha_4=0$, and $\alpha_2+\alpha_5=0$, on may then write
$L(i\omega_e,p_1^0;\alpha)=\Lambda(p_1^0+i\alpha_1,-i\omega_e)$.

\begin{figure}[htbp]
\begin{center}
\includegraphics[width=5cm]{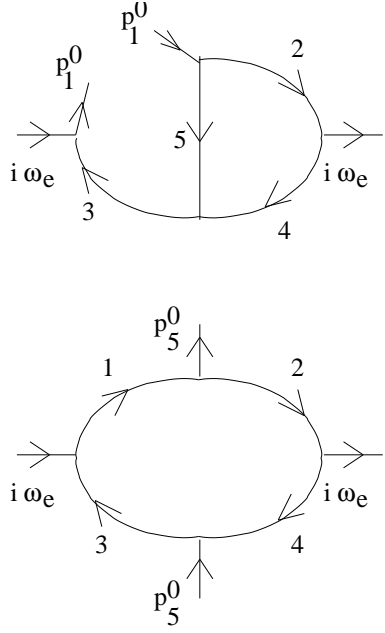}
\caption{Diagrammatic representations of the contributions to 
$I_1(i\omega_e,p_1^0;\alpha)$ (top, see Eq.~(\ref{I1analytic})), and
$I_5(i\omega_e,p_5^0;\alpha)$ (bottom, see Eq.~(\ref{eq:den1})).\label{fig2a}}
\end{center}
\end{figure}

One can  treat $I_2$, $I_3$ and $I_4$ in the same way (one obtains the further conditions $\alpha_4+\alpha_5=0$, $\alpha_5-\alpha_1=0$, $\alpha_5-\alpha_3=0$, $\alpha_1-\alpha_3=0$ which are compatible with the previous ones). For $I_5$, 
corresponding to the bottom diagram in
Fig.~\ref{fig2a},  a  similar analysis
   can be carried out.  One has:
\begin{eqnarray}\label{eq:den1}
I_5\left\{i\omega_e,p^0_5;\alpha\right\}  =  \int_{1234} & & \left\{
\frac{-\theta(-p^0_1)}{(p^0_3-p^0_1+i\omega_e)(p^0_2+p^0_5-p^0_1)(p^0_4+p^0_5-p^0_1+i\omega_e)}\right.\nonumber\\ 
& + &
\frac{-\theta(-p^0_3)}{(p^0_1-p^0_3-i\omega_e)(p^0_4+p^0_5-p^0_3)(p^0_2-p^0_3+p^0_5-i\omega_e)}\nonumber\\ 
& + &
\frac{-\theta(-p^0_2)}{(p^0_3-p^0_2-p^0_5+i\omega_e)(p^0_1-p^0_2-p^0_5)(p^0_4-p^0_2+i\omega_e)}\nonumber\\ 
& + &
\left.\frac{-\theta(-p^0_4)}{(p^0_1-p^0_4-p^0_5-i\omega_e)(p^0_3-p^0_4-p^0_5)(p^0_2-p^0_4-i\omega_e)}
\right\}, \nonumber\\
\end{eqnarray}
where again it is understood that all the variables $p_1,p_2,p_3,p_4$ have a
small imaginary part.
One considers then the 4-point function associated to the bottom 
diagram in Fig. \ref{fig2a}, calculated at
zero temperature. By applying the rules of Sect.~\ref{rules}  one obtains
\begin{eqnarray}\label{eq:den1a}
J_5\left\{i\omega_a,i\omega_b,i\omega_c\right\}  =  \int_{1234} & & \left\{
\frac{-\theta(-p^0_1)}{(p^0_3-p^0_1+i\omega_a)(p^0_2
-p^0_1-i\omega_b)(p^0_4-p^0_1-i\omega_b-i\omega_c)}\right.\nonumber\\
& + &
\frac{-\theta(-p^0_3)}{(p^0_1-p^0_3-i\omega_a)(p^0_4
-p^0_3-i\omega_a-i\omega_b-i\omega_c)(p^0_2-p^0_3-i\omega_a-i\omega_b)}\nonumber\\ 
& + &
\frac{-\theta(-p^0_2)}{(p^0_3-p^0_2+i\omega_a+i\omega_b)(p^0_1-p^0_2+i\omega_b)(p^0_4-p^0_2-i\omega_c)}\nonumber\\ 
& + &
\left.\frac{-\theta(-p^0_4)}{(p^0_1-p^0_4+i\omega_b+i\omega_c)(p^0_3-p^0_4+i\omega_a+i\omega_b+i\omega_c)(p^0_2-p^0_4+i\omega_c)}
\right\}.\nonumber\\
\end{eqnarray}
Again it is possible to find a set of regulators and a domain of 
analyticity of $J_5$ where $J_5$ and the
regulated integral
$I_5$ coincide.  By comparing Eqs.~(\ref{eq:den1}) and 
(\ref{eq:den1a}), one finds the following relations:
\beq
i\omega_a=i\omega_e-i\alpha_1+i\alpha_3\qquad
i\omega_b=-p_5^0+i\alpha_1-i\alpha_2-i\alpha_5\qquad
i\omega_c=-i\omega_e+i\alpha_2-i\alpha_4.
\eeq
These allow us to write
\beq
I_5(i\omega_e,p_5^0;\alpha)= J_5(i\omega_e,-p_5^0-i\alpha_5,-i\omega_e),
\eeq
where the constraints discussed before Eq.~(\ref{Gaudin}) can be 
satisfied with the choice
$\alpha_2=\alpha_4$, $\alpha_1=\alpha_3$, and $\alpha_5\ne 0$ (note that $\alpha_2=-\alpha_1$ in order for these conditions to be compatible with those derive for $I_1$, $I_2$, $I_3$ and $I_4$).

\begin{figure}[htbp]
\begin{center}
\includegraphics[width=5cm]{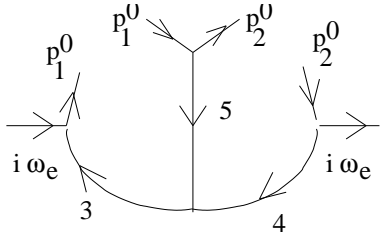}
\caption{Diagrammatic representation of the contribution 
$I_{12}(i\omega_e,p^0_1,p^0_2;\alpha)$ in 
Eq.~(\ref{I12}).\label{fig3a}}
\end{center}
\end{figure}

Finally, for completeness, we consider the terms with two factors 
$\sigma$ (single blocs in
Fig. \ref{fig4}), which raise in fact no real problems. Using the 
spectral representation for the
propagator we perform all the integrals over frequencies which are
not involved in $\sigma$. The resulting contribution takes the form 
of a product of propagators. For instance,
$I_{12}$ can be written as  follows:
\begin{equation}\label{I12}
I_{12}(i\omega_e,p^0_1,p^0_2;\alpha)=D(p^0_1-i\omega_e)D(p^0_1-p^0_2-i\alpha_5)D(p^0_2-i\omega_e),
\end{equation}
giving the following contribution to $I\{\omega_e\}$:
\beq
\int\frac{dp^0_1}{2\pi}\frac{dp^0_2}{2\pi}\sigma(p^0_1)\sigma(p^0_2)I_{12}
(i\omega_e,p^0_1,p^0_2;\alpha).
\eeq
The subdiagram corresponding to $I_{12}$
  is represented in Fig. \ref{fig3a}.
The expression (\ref{I12}) is well defined for all values of 
$p^0_1,p^0_2$,  thanks to the
regularization.

In this subsection, we have achieved our goal of expressing all the 
vacuum subdiagrams of the two-loop diagram of Fig.~\ref{fig1}
in terms of analytic continuation of vacuum amplitudes (here 3-point 
and 4-point functions, or isolated propagators).
  Before going any further, we explain how this can be used in a simple analysis
  of ultraviolet divergences of Feynman diagrams at finite temperature.

\subsection{Eliminating ultraviolet divergences in  the two-loop example}

\begin{figure}[htbp]
\begin{center}
\includegraphics[width=7cm]{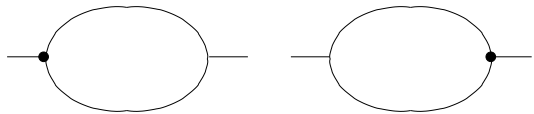}
\caption{Counter-terms diagrams; in each diagram, one vertex is 
associated to a coupling constant $g$,
the other to the coupling constant counterterm $\delta g$.\label{fig:2}}
\end{center}
\end{figure}

We consider again the two-loop diagram of Fig.~\ref{fig1}, calculated with the
propagator $D_0$. It is  of order $g^4$, where $g$ is the coupling 
constant. We focus on the contributions
involving an ultraviolet divergent subdiagram inside a finite 
temperature integral. These are the
contributions with one  thermal factor. By applying  the rules of 
Sect.~\ref{rules}, we obtain
\begin{equation}\label{I1div0}
I^{(1)}(i\omega_e;\alpha)=2g^4\int_{P_1}\sigma(p_1^0)I_1(i\omega_e,p_1^0;\alpha)+\frac{g^4}{2}
\int_{P_5}\sigma(p_5^0)I_5(i\omega_e,p_5^0;\alpha),
\end{equation}
where the notation $\int_P$ is for an integral over the four 
components of the momentum $P$.
The factor $1/2$ in the second term is the symmetry factor; the 
factor 2 in the first term arises from the  4
contributions identical to that of the first block in  Fig. 
\ref{fig:2}. Note that only the first integral
contains ultraviolet divergences ($I_1(i\omega_e,p_1^0;\alpha)$ is ultraviolet 
divergent).

We shall show that such temperature dependent, ultraviolet divergent, 
contributions are cancelled by the
counterterm of order $g^3$ which eliminates the divergence in the 
one-loop contribution to the three-point
vertex (see Fig.~\ref{fig:4}). Consider then the diagrams of 
Fig.~\ref{fig:2} computed at finite
temperature. We get:
\begin{equation}
I_{ct}(i\omega_e;\alpha)=2g\delta 
g\int_{P_1}\sigma(p_1^0)D_0(p_1^0+i\alpha_1-i\omega_e),
\end{equation}
where $\delta g$ is the counterterm which eliminates the divergence 
of the diagram of Fig.~\ref{fig:4}.
Combining this term with the first term of Eq.~(\ref{I1div0}), one gets
\begin{equation}
2g\int_{P_1}\sigma(p_1^0)D_0(p_1^0+i\alpha_1-i\omega_e)\left[L(i\omega_e,p_1^0;\alpha)+\delta g\right].
\end{equation}
We argue now that $\left[L(i\omega_e,p_1^0;\alpha)+\delta
g\right]$ is finite. This follows from the fact that 
$L(i\omega_e,p_1^0;\alpha)$ is the analytic continuation
of the three-point vacuum amplitude $\Lambda(P_a,P_b)$ defined in 
Eq.~(\ref{Lambda0}), and  the divergence of
this amplitude is precisely that which is cancelled by the 
counterterm $\delta g$.
\begin{figure}
\begin{center}
\includegraphics[width=7cm]{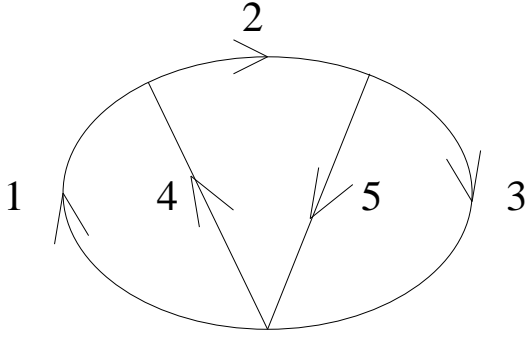}
\caption{Counter-example\label{fig:3loop}}
\end{center}
\end{figure}
\subsection{Counter-example}
We now present an example of a diagram which contains vacuum 
subdiagrams that cannot be simply related
  to analytic continations of the topologically equivalent vacuum 
amplitudes. This diagram is displayed
in Fig. \ref{fig:3loop}.

We start by applying the rules of Sect.~\ref{rules} and obtain:
\beq
I=\int_{12345}\hat I,
\eeq
with
\begin{eqnarray}\label{Ihat}
\hat I &= &
-\frac{n_{2}n_{4}n_{5}+n_{2}n_{-1}n_{-3}+n_{2}n_{-1}n_{5}+n_{2}n_{4}n_{-3}}{(p^0_2-p^0_1-p^0_4)(p^0_3+p^0_5-p^0_2)}\nonumber\\
&  & \hspace{0.05cm}
-\frac{n_{3}n_{5}n_{4}+n_{3}n_{5}n_{-1}}{(p^0_1+p^0_4-p^0_3-p^0_5)(p^0_3+p^0_5-p^0_2)}\nonumber\\
& & \hspace{0.05cm} -\frac{n_{1}n_{4}n_{5}+n_{1}n_{4}n_{-3}}{(p^0_1+p^0_4-p^0_3-p^0_5)(p^0_2-p^0_1-p^0_4)},
\end{eqnarray}
where we have used the shorthand notation introduced after 
Eq.~(\ref{eq:Gaudin_ex}).

Let us first verify that this formula has no infrared singularity 
associated with vanishing
denominators. To this aim, let us set
\beq
A=p_2^0\, ,\qquad B=p_3^0+p_5^0\, ,\qquad C=p_1^0+p_4^0\, .
\eeq
Then one can rewrite $\hat I$ as follows:
\beq
\hat I=(n_4+n_{-1})(n_5+n_{-3})
\left\{
\frac{n_A}{(A-B)(A-C)}+\frac{n_B}{(B-A)(B-C)}+\frac{n_C}{(C-A)(C-B)}
\right\},\nonumber\\
\eeq
where we have used relations such as $n_1n_4=n_{1+4}(n_4+n_{-1})$. 
Consider now what happens when $A-B\to 0$
with $C\ne B$. Then, the first two terms in the brackets, wich are 
potentially divergent, lead  in fact to a
well defined limit:
\beq
\frac{n_A}{(A-B)(A-C)}+\frac{n_B}{(B-A)(B-C)}\to \frac{n'_B 
(B-C)-n_B}{(B-C)^2},
\eeq
where $n'_B$ denotes the derivative of $n_B$ with respect to $B$. If 
we further let $B-C\to 0$, we get
\beq
\frac{n'_B(B-C)-n_B+n_C}{(B-C)^2}\rightarrow \frac{1}{2} n''_C,
\eeq
where $n''_C$ is the second derivative of $n_C$ with respect to $C$. 
Thus the limits where denominators vanish is well defined.

Now, in order to manipulate freely the various terms of 
Eq.~(\ref{Ihat}), we introduce an infrared regularization  by 
shifting   $p_j^0$ by a small
imaginary part, $p_j^0\to  p_j^0+i\alpha_j$. The various $\alpha_j$ 
thus introduced are
not all independent. Indeed,  one may identify only three different 
factors in the denominators:
\begin{eqnarray}
a = p^0_1+p^0_4-p^0_3-p^0_5\, ,\qquad
b=p^0_2-p^0_1-p^0_4\, ,\qquad
c = p^0_3+p^0_5-p^0_2\, ,
\end{eqnarray}
which furthermore satisfy the relation
\begin{equation}
a+b+c=0.
\end{equation}
Thus there are only two independent imaginary parts that we can play 
with. We shall  attribute small
imaginary parts to each of the factors $a$, $b$ and $c$, 
respectively $\alpha_a$, $\alpha_b$ and
$\alpha_c$, with the following constraints:
\begin{eqnarray}
  \alpha_a\neq 0\,,\,\alpha_b\neq 0\,,\,\alpha_c\neq 0, \qquad
  \alpha_a+\alpha_b+\alpha_c=0,
\end{eqnarray}
from which it follows that:
\begin{eqnarray}\label{constraints2}
\alpha_a+\alpha_b\neq 0,\qquad
\alpha_b+\alpha_c\neq 0,\qquad
\alpha_a+\alpha_c\neq 0.
\end{eqnarray}
The regularized expression of $\hat I$ in Eq.~(\ref{Ihat}) reads then:
\begin{eqnarray}
\hat I & = &
-\frac{n_{2}n_{4}n_{5}+n_{2}n_{-1}n_{-3}+n_{2}n_{-1}n_{5}+n_{2}n_{4}n_{-3}}{(p^0_2-p^0_1-p^0_4+i\alpha_b)(p^0_3+p^0_5-p^0_2+i\alpha_c)}\nonumber\\
&  & \hspace{0.05cm}
-\frac{n_{3}n_{5}n_{4}+n_{3}n_{5}n_{-1}}{(p^0_1+p^0_4-p^0_3-p^0_5+i\alpha_a)(p^0_3+p^0_5-p^0_2+i\alpha_c)}\nonumber\\
& &
-\frac{n_{1}n_{4}n_{5}+n_{1}n_{4}n_{-3}}{(p^0_1+p^0_4-p^0_3-p^0_5+i\alpha_a)(p^0_2-p^0_1-p^0_4+i\alpha_b)}.\nonumber\\
\end{eqnarray}
It has a well defined limit when $\alpha\to 0$.

We now split each statistical factor into a vacuum and a thermal 
piece and proceed to the identification of
the various subdiagrams. We can go through  the same
analysis as before, and identify vacuum amplitudes,  except for the 
terms with one thermal factor.
Consider in particular the contribution $I_1$, the contribution where 
the line 1 is a thermal line. It is
given by
\begin{eqnarray}
\hat I_1(p_1^0;\alpha)& = &
-\frac{\theta(p^0_2)\theta(p^0_3)+\theta(-p^0_2)\theta(-p^0_5)}{(p^0_2-p^0_1-p^0_4+i\alpha_b)(p^0_3+p^0_5-p^0_2+i\alpha_c)}\nonumber\\
&  & \hspace{0.05cm}
-\frac{\theta(-p^0_3)\theta(-p^0_5)}{(p^0_1+p^0_4-p^0_3-p^0_5+i\alpha_a)(p^0_3+p^0_5-p^0_2+i\alpha_c)}\nonumber\\
& &
-\frac{\theta(-p^0_4)\theta(-p^0_5)-\theta(-p^0_4)\theta(p^0_3)}{(p^0_1+p^0_4-p^0_3-p^0_5+i\alpha_a)(p^0_2-p^0_1-p^0_4+i\alpha_b)}.\nonumber\\
\end{eqnarray}
One sees that in the denominators $p^0_1$ appears in the combinations 
$p^0_1+i\alpha_a$ or $p^0_1-i\alpha_b$. If one whishes to regard the 
function
as an analytic continuation of a 2-point function depending on a 
single variable, it is then necessary to have $\alpha_a=-\alpha_b$. 
But this is in
contradiction with the constraints (\ref{constraints2}).  The other 
integrals  $I_2$, $I_3$, $I_4$, $I_5$
suffer from the same difficulty.

\section{Conclusions\label{sec:conclu}}

Gaudin's method to perform the sums over the Matsubara frequencies 
leads to a very simple scheme for
calculating Feynman diagrams at finite temperature: Once one has
identified all  the trees in the diagram, one gets for each tree a 
contribution in the form of a fraction
whose numerator and  denominator are given by simple rules. This 
method enables one to establish  general properties such as those 
treated in the
appendix. It can also be easily implemented on a computer \cite{urko}.

  One of our main concerns in this paper was the identification of 
vacuum subdiagrams, and their possible relations
  to analytical continutations of Euclidean amplitudes. This is 
relevant in particular to the discussion of ultraviolet divergent
  contributions in calculations at finite temperature. Gaudin's 
formula is useful in this context as it leads automatically to 
expressions in which
the arguments of the  statistical factors are the frequencies 
attached to the lines of the diagram, rather than combinations
of such frequencies. This makes it easy to separate, in the total 
contribution of
the diagram, thermal
     parts (sets of lines corresponding to the statistical factors), 
and vacuum parts (the rest of
the lines in the diagram). Since the temperature
      cuts the flow of momentum in the thermal lines, the ultraviolet 
behaviour arises from the vacuum parts
(note that this is true for generic propagators, not only for the 
perturbative one; that is,
  this reasoning applies to the general discussion in 
\cite{Blaizot:2003an}). In several cases of practical
interest, we have been able to relate these vacuum parts
      to vacuum amplitudes. However for general diagrams this 
identification is not
always possible, at least in the way we have followed. The difficulty 
arises from the necessity to
introduce a regularization which gives a meaning to isolated terms in 
Gaudin's formula. For the
   regularization that we have studied, we have shown that it is not 
always possible to identify
vacuum parts with vacuum amplitudes. (Note that this difficulty does 
not alter the general proof
given in \cite{Blaizot:2003an}; it only makes its practical 
implementation more difficult.)

Finally, we note that the generalization to theories other than 
scalar theories is straightforward, since all the
information about the theory is encoded in the spectral function 
which does not need to be specified in most part of
  the analysis presented in this paper.

\section*{Acknowledgment} The authors would like to thank E. Iancu for fruitful discussions.

\appendix



\section{Proof of the conjecture of Ref.~\cite{Esp:2003} }

In order to illustrate the power of
Gaudin's technique to calculate sums over Matsubara frequencies,  we 
give here a simple proof of the main part of the
conjecture stated in
\cite{Esp:2003}. This conjecture concerns the possibility to 
reconstruct the expression of a Feynman
diagram at finite temperature, starting from its corresponding 
expression at zero temperature, in the
imaginary time formalism.  The authors of Ref.~\cite{Esp:2003} 
express their conjecture in an algebraic
way, using a ``thermal operator''. Here we shall only show how the 
algorithm underlying their result emerges
naturally from the rules of Sect.~\ref{rules}.

Let us  consider first the one-loop example of section II. We have 
obtained its expression at finite
temperature, Eq.~(\ref{oneloop1}), which we recall here for convenience:
\begin{eqnarray}\label{ex}
I(i\omega_e,{\bf k}) =
\int\frac{d^3p}{(2\pi)^3}\frac{1}{2\varepsilon_p}\frac{1}{2\varepsilon_q}
&\!&\!\!\!\!\!\left\{(1+n_{\varepsilon_p}+n_{\varepsilon_{q}})
\left(\frac{1}{i\omega_e+\varepsilon_p+\varepsilon_{q}}-\frac{1}{i\omega_e-\varepsilon_p-\varepsilon_{q}}\right)\right.\nonumber\\
&\,&+\left.
(n_{\varepsilon_p}-n_{\varepsilon_{q}})
\left(\frac{1}{i\omega_e-\varepsilon_p+\varepsilon_{q}}-\frac{1}{i\omega_e+\varepsilon_p-\varepsilon_{q}}\right)\right\},\nonumber\\
\end{eqnarray}
with ${\bf q}={\bf k}-{\bf p}$. The vacuum result is obtained from 
(\ref{ex}) by
dropping the terms proportional to the statistical factors 
($n_{\varepsilon_p}=0$ at zero
temperature):
\begin{equation}\label{ex0}
I^{(0)}(i\omega_e,{\bf k}) =
\int\frac{d^3p}{(2\pi)^3}\frac{1}{2\varepsilon_p}\frac{1}{2\varepsilon_q}\left\{\frac{1}{i\omega_e+\varepsilon_p+\varepsilon_{q}}-\frac{1}{i\omega_e-\varepsilon_p-\varepsilon_{q}}\right\}.
\end{equation}
The finite temperature contribution, $I^{(1)}$,  is the sum of terms 
proportional to the statistical factors.
The authors of \cite{Esp:2003} propose a simple algorithm to 
reconstruct  $I^{(1)}$   from $I^{(0)}$. For
each energy $\varepsilon_p$ or $\varepsilon_q$  appearing in the 
denominators of $I^{(0)}$, one adds a term
proportional to a statistical  factor
$n_{\epsilon_p}$ or $n_{\epsilon_q}$ multiplied by a sum of two
energy denominators, one of which is the original denominator of 
$I^{(0)}$, the other being obtained from it
through the replacement
$\varepsilon_p\rightarrow  -\varepsilon_p$ or 
$\varepsilon_q\rightarrow -\varepsilon_q$:
\begin{eqnarray}
\frac{1}{i\omega_e+  \varepsilon_p +\varepsilon_{q}}  &\rightarrow &
  n_{\epsilon_p}\left\{\frac{1}{i\omega_e+
\varepsilon_p +\varepsilon_{q}}+\frac{1}{i\omega_e
-\varepsilon_p +\varepsilon_{q}}\right\}\nonumber\\
&+&   n_{\epsilon_q} \left\{\frac{1}{i\omega_e+
\varepsilon_p +\varepsilon_{q}}+\frac{1}{i\omega_e+\varepsilon_p
-\varepsilon_{q} }\right\},\nonumber\\
\frac{1}{i\omega_e-  \varepsilon_p -\varepsilon_{q}} & \rightarrow
&   n_{\epsilon_p} \left\{\frac{1}{i\omega_e-
\varepsilon_p -\varepsilon_{q}}+\frac{1}{i\omega_e
+\varepsilon_p -\varepsilon_{q}}\right\}\nonumber\\
  & +
&  n_{\epsilon_q} \left\{\frac{1}{i\omega_e-
\varepsilon_p -\varepsilon_{q}}+\frac{1}{i\omega_e-\varepsilon_p
+\varepsilon_{q} }\right\}.
\end{eqnarray}
This procedure emerges naturally if one writes $I$ as follows:
\begin{equation}\label{Gaudin_eye}
I(i\omega_e,k)=\int_{-\infty}^{\infty}\frac{dp_0}{2\pi}\frac{dq_0}{2\pi}\rho_0(p_0)\rho_0(q_0)\frac{n_{p_0}-n_{-q_0}}{p_0+q_0-i\omega_e},
\end{equation}
with the free spectral density given by:
\begin{equation}
\rho_0(p_0)=2\pi\epsilon(p_0)\delta(p_0^2-\varepsilon_p^2)=\frac{\pi}{\varepsilon_p}\left\{\delta(p_0-\varepsilon_p)-\delta(p_0+\varepsilon_p)\right\}.
\end{equation}
Indeed each statistical factor $n_{p_0}$ contains a vacuum part
$-\theta(-p_0)$ which selects one of the two peaks in the
spectral density, giving in both cases a positive contribution,  and 
a thermal part $\epsilon_{p_0}n_{|p_0|}$
with which both peaks in the spectral density contribute an equal and 
positive amount.

This result is easily generalized, as we show now. Let us  consider a 
general diagram of perturbation
theory  ($\rho=\rho_0$) for which the sum over Matsubara frequency 
leads to the following integral (see
Eq.~(\ref{Gaudin})):
\begin{equation}\label{Gaudin_again}
I\left\{i\omega_e\right\}=\sum_{{\cal T}}\prod_{l\in
\bar{{\cal T}}}\int\frac{dp^0_l}{(2\pi)}\rho(p_{l}^0)\epsilon_l
n_{\epsilon_l p_l^0}\prod_{j\in
{\cal T}}D (\Omega_j;\alpha),
\end{equation}
where $\alpha$ denotes the regulators (see Sect.~\ref{section:reg}). 
We shall focus on the contribution of a
given tree, denoted by $I(\mathcal{T};\alpha)$. This  contains a
  vacuum contribution obtained after replacing each line of $\bar 
{\cal T}$ by a vacuum line  carrying a factor
  $-\epsilon_l
\theta(-{\epsilon_l p_{l}^0})$. We shall denote this contribution by
$I^{(0)}(\mathcal{T};\alpha)$:
\begin{equation}\label{IS0}
I^{(0)}(\mathcal{T};\alpha)=\prod_{l\in
\bar{{\cal
T}}}\int\frac{dp^0_l}{(2\pi)}\rho(p_{l}^0)\left\{-\epsilon_l
\theta(\epsilon_l p_l^0)\right\}\prod_{j\in
{\cal T}}D (\Omega_j;\alpha).
\end{equation}
   The other contributions to $I(\mathcal{T};\alpha)$ are obtained by 
replacing some of the vacumm
lines in $I^{(0)}(\mathcal{T};\alpha)$ by thermal lines carrying 
factors $\varepsilon(p_{l}^0)n_{|p_{l}^0|}$.
The contribution for which the subset $\mathcal{S}$ of lines of 
$\bar{\mathcal{T}}$ are thermal lines reads:
\begin{equation}\label{IS}
I_{\mathcal{S}}(\mathcal{T};\alpha)=\int\prod_{{\bar l}\in
\bar{{\cal
T}}}\frac{dp^0_{\bar l}}{(2\pi)}\rho(p_{{\bar l}}^0)\prod_{l'\in\bar{\cal
S}}\left\{-\epsilon_{l'} \theta(-{\epsilon_{l'}
p_{l'}^0})\right\}\prod_{l\in{\cal S}}\epsilon_l n_{\epsilon_l
p_{l}^0}\prod_{j\in{\cal T}}D (\Omega_j;\alpha),
\end{equation}
where $\bar{\cal S}$ denotes the set of lines of $\bar{\cal T}$ which 
remain vacuum lines.

At this point we can repeat the same argument as in the one-loop 
example above. In
$I^{(0)}(\mathcal{T};\alpha)$, when we integrate over the free spectral
densities, each $\theta$-function selects one of the peaks in the
spectral density and gives always a positive contribution, whatever 
the selected peak is.  The denominators
are determined by plugging in the
$\Omega_j$'s the energies $\varepsilon_p$ corresponding to the 
(selected) peaks of the spectral functions,
with the signs given by the rules of Sect.~\ref{rules}. In going from 
$I^{(0)}(\mathcal{T};\alpha)$ to
$I_{\mathcal{S}}(\mathcal{T};\alpha)$, we replace some of
$\theta$-functions   by  contributions that are proportional to a 
thermal factor and in which the two peaks
in the spectral density contribute on the same footing, leading to a 
duplication of denominators
with the values $\pm\varepsilon_p$ of the energies. This is 
essentially the content of ``Statement 1'' in \cite{Esp:2003},
the thermal operator introduced there being the operator relating 
$I_{\mathcal{S}}$ in Eq.~(\ref{IS}) to $I^{(0)}$ in Eq.~(\ref{IS0}).

\end{document}